\documentclass{emulateapj} 
\usepackage{enumerate,multirow}
\usepackage{amsmath,amssymb,graphicx,natbib,subfigure,dcolumn,bm,comment} 
\usepackage{amsmath}
\usepackage{amssymb} 
\usepackage{color}

\shorttitle{Triples inside globular clusters}
\shortauthors{Antonini et al.}

\begin{document}
\def\gap{\;\rlap{\lower 2.5pt
\hbox{$\sim$}}\raise 1.5pt\hbox{$>$}\;}
\def\lap{\;\rlap{\lower 2.5pt
 \hbox{$\sim$}}\raise 1.5pt\hbox{$<$}\;}

\def\eb#1{{\textcolor{blue}{[[\bf EB: #1]]}}}
\def\mb#1{{\textcolor{red}{[[\bf MB: #1]]}}}

\newcommand\MBH{\rm MBH} 
\newcommand\NC{NSC}
\newcommand\CMO{CMO} 
\title{black hole mergers and blue stragglers from hierarchical triples formed in globular clusters}

\author{Fabio Antonini \\  Sourav Chatterjee,
Carl~ L. Rodriguez,  Meagan Morscher, Bharath Pattabiraman \\   
Vicky Kalogera, Frederic A. Rasio}
\affil{Center for Interdisciplinary Exploration and Research in Astrophysics (CIERA)
and Department of Physics and Astrophysics, 
Northwestern University,  Evanston, IL 60208}

\begin{abstract}
Hierarchical triple-star systems are expected to form frequently via close binary-binary 
encounters in the dense cores of globular clusters.
In a sufficiently inclined triple, gravitational interactions         
between the inner and outer binary can cause large-amplitude oscillations in the eccentricity of
the inner orbit\ (``Lidov-Kozai cycles''), which can lead to a collision and merger of the two inner components.
In this paper we use Monte Carlo models of dense star clusters to identify all triple systems formed dynamically
and we compute their evolution using a highly accurate
three-body integrator which incorporates relativistic and tidal effects.
We find that a large fraction of these triples evolve through a non-secular 
dynamical phase which can drive the inner binary to higher eccentricities than 
predicted by the standard secular perturbation theory (even including octupole-order terms).
We place constraints on the importance of Lidov-Kozai-induced mergers for producing: 
(i) gravitational wave sources detectable by Advanced LIGO\ (aLIGO), for triples with an inner pair of 
stellar black holes;
and (ii) blue straggler stars, for triples with main-sequence-star components.
We find a realistic  aLIGO detection rate of black hole mergers due to the Lidov-Kozai mechanism 
of $\sim 1\ \rm yr^{-1}$, with about $20\%$ of these having a finite eccentricity 
 when they first chirp into the aLIGO frequency band. While rare, these events are likely to 
dominate among eccentric compact object inspirals that are potentially detectable by aLIGO.
For blue stragglers, we find that the Lidov-Kozai mechanism
can contribute up to $\sim 10\% $ of their total numbers in globular clusters.
\end{abstract}

\section{Introduction}
Observations indicate that a large fraction of  stars
reside in triple systems\ \citep[][and references therein]{2015ApJ...799....4R}.
In a dynamically stable non-coplanar triple the gravitational interaction between the inner and outer
binary drives periodic/quasi-periodic variations
of the mutual inclination between the two orbits
and of the eccentricity of the inner binary\ \citep{1961Lidov,1962AJ.....67..591K}.
Such eccentricity oscillations, first discovered by \citet{1961Lidov},
can lead to a close approach between the inner binary components which can produce 
a variety of exotic astrophysical phenomena, including compact-object and stellar 
binary mergers\ \citep[e.g.,][]{2000ApJ...535..385F,2002ApJ...576..894M,2010ApJ...713...90A,2014ApJ...793..137N},
X-ray binaries, type Ia supernovae and Gamma-ray bursts\ \citep{2011ApJ...741...82T,2013MNRAS.430.2262H}.

Lidov-Kozai\ (LK) oscillations are usually described as a secular phenomenon, occurring on
timescales much longer than the orbital period of the 
outer binary orbit\ \citep[e.g.,][]{2000ApJ...535..385F,2007ApJ...669.1298F,2013MNRAS.431.2155N}. 
The standard approach is to make use of  the so-called  orbit-averaging technique in which
one computes the mutual gravitational interaction\ (torques) between two mass weighted ellipses 
instead of point masses on orbits. The basic assumption is that the evolution 
of the orbital parameters occur on a timescale 
much longer than the dynamical timescales of the system\ \citep[e.g.,][]{2013degn.book.....M}.

The orbit averaged equations of motion
can describe the evolution of a wide range of systems, but 
become inaccurate once the triple hierarchy is moderate.
\citet{2012ApJ...757...27A}  showed that the orbit-average treatment, even at the octupole level
of approximation, becomes inaccurate in hierarchical systems
where the tertiary comes closer to the inner binary than a certain distance.
At the high eccentricity phase of a LK cycle 
the timescale for the angular momentum  perturbation due to the tertiary
can become comparable to or shorter than the dynamical timescales of the system rendering 
 the secular approach 
no longer valid\ 
\citep{2012ApJ...757...27A,2012arXiv1211.4584K,2013PhRvL.111f1106S,2014ApJ...781...45A,2014MNRAS.439.1079A,2014MNRAS.438..573B}.
In this case, the binary can be driven to a much closer separation than
the secular theory would otherwise predict. This can have important consequences for
 the evolution of the inner binary.
For example, \citet{2012ApJ...757...27A},  \citet{2013PhRvL.111f1106S} and 
\citet{2014ApJ...781...45A}  demonstrated that in this non-secular regime,
a compact object binary can be pushed 
inside the $10\ $Hz gravitational wave\ (GW) frequency band of
Advanced Ligo\ (aLIGO) detectors with a finite eccentricity. 
Similarly, in the case of stellar binaries, the two stars
could be driven to a direct collision. \citet{2014ApJ...793..137N} and  \citet{2015ebss.book..251P} suggested that
direct stellar collisions driven by the LK mechanism  might have 
an important role in the formation of blue straggler stars\ (BSSs) in stellar clusters.

Globular clusters\ (GCs)  are a natural environment 
for forming triples and higher order multiples.
On one hand,  the dense stellar environment of GCs 
is hostile to triples. In fact,
given the high densities of GCs, 
triples may be involved in close encounters which tend to destroy them.
On the other hand, 
the high stellar densities that characterize the core of GCs favor dynamical binary-binary
interactions that  naturally lead to the formation of stable triples\  \citep{1993Natur.364..423S},
including black hole\ (BH) triples and mixed, compact object-star, triples that could not otherwise form. 
The inner binaries in the  triples may be driven to merge before their next interaction.
Interestingly, such mergers have been suggested
 as a channel for the formation of 
compact-object mergers\ \citep{2002ApJ...576..894M,2003ApJ...598..419W,2016MNRAS.455...35A}, 
and more recently  as a channel for the formation of BSSs through 
mass transfer and merger\ \citep{2009ApJ...697.1048P,2011MNRAS.410.2370L,2013AJ....145....8G,2014ApJ...793..137N,2015ebss.book..251P}.

In this paper we study the evolution of  dynamically assembled triples inside GCs
with particular focus on triples containing  an inner main-sequence\ (MS) stellar binary  and triples with an inner BH binary. 
We do not consider here the evolution of mixed star-compact object systems, or binaries 
containing evolved (e.g., red giant) stars; we reserve a study of these latter systems to a future investigation. 
In this paper we use a Monte Carlo code to simulate the long-term evolution of GCs\ \citep{2015ApJ...800....9M},
and identify all triple systems formed during the cluster evolution.
We then evolved these triples using a high-accuracy direct three-body integrator which includes
tidal and general relativity (GR) effects for both the inner and
outer orbit.  Scattering with
other stars is also accounted for, but only in the sense that
 the triples are evolved for a time equal to the timescale over which
they will be disrupted  through gravitational  encounters
with other stars. The results of the direct integrations are used 
to determine the rate of BH and stellar binary mergers induced by 
the LK mechanism  in GCs.

Although previous studies have shown that triples can form quite frequently in GCs, their detailed long-term 
dynamical evolution in the cluster environment  has not been yet investigated.
Our study sheds light on exactly this
point, i.e. it allows us to assess for the first time the role of the host cluster evolution and its properties 
on the formation and dynamics of the triples.  This allows us to place reliable 
constraints on the contribution
of mergers from triples to BSS populations and on the event rate and properties of inspiraling  BH binaries
as a source of GW radiation for aLIGO detectors\ \citep[e.g.][]{2015CQGra..32k5012A,2015CQGra..32b4001A}.

We note  that the  dynamics  of  triples with either an inner stellar or compact object component
has  been investigated  by many   authors  and in  a variety of 
contexts \   (including stellar triples, planets around stars, 
supermassive BH mergers),
using   the   standard  orbit averaged formalism and therefore fully neglecting non-secular effects\ 
\citep[e.g.,][]{2002ApJ...578..775B,2003ApJ...598..419W,2011Natur.473..187N,2013MNRAS.431.2155N,2013MNRAS.430.2262H,2014ApJ...793..137N}.
The new ingredients of our study compared to previous work
dealing with similar problems are:
(i)  the use of realistic initial conditions for our triple populations, 
 that were directly obtained using
 Monte Carlo models of the long-term dynamical evolution of 
 GCs; (ii) the use of accurate three-body integrations that allows
 an essentially exact treatment of 
 the triple dynamics, avoiding the approximations that are made when using 
the secular equations of motion. 

The paper is organized as follows.
We begin with a discussion of dynamically assembled triples in GCs and their properties 
in Section\ \ref{bd-oaa}, and show that for most  triples the orbit average approximation  
is expected to break down.
In Section\ \ref{meth+ICs} we discuss our GC models, triple initial conditions and adopted numerical methods.
We present our results on BH-triples and associated aLIGO event rates in
Section\ \ref{res-bhs}.
In Section\ \ref{st-tr} we describe the results concerning stellar triples and their 
implications for BSS formation in GCs. 
We further discuss the implications of our results in Section\ \ref{disc}. Section\ \ref{summ} sums up. 

\section{break-down of the secular approximation}\label{bd-oaa}
We consider a binary of components of masses $m_{0}$
and $m_{1}$ orbiting a third  body of mass $m_2$.
We indicate the eccentricities of the inner and outer orbits, respectively,
as $e_{1}$ and $e_{2}$, and semi-major axes $a_{1}$ and $a_{2}$; we 
define $\omega_{1}$ as the argument of periapsis of the inner
binary relative to the line of the descending node, $\omega_2$ as the argument of periapsis 
of the outer orbit, and $I$ as the
orbital inclination of the inner orbit relative to the outer orbit.

In what follows  we only consider  triple
systems that satisfy the stability criterion of~\citet{2001MNRAS.321..398M},
\begin{eqnarray}\label{stab}
\frac{a_2}{a_1} &>& \frac{3.3}{1-e_2}\left[ \frac{2}{3}\left(1+\frac{m_2}{M_b}\right) 
\frac{1+e_2}{\left(1-e_2\right)^{1/2}}\right]^{2/5}\\
&&\times(1-0.3I/\pi)~, \nonumber
\end{eqnarray}
with $M_{b}=m_{0}+m_{1}$.

If the system satisfies the stability criterion~(\ref{stab}) it is also reasonable to describe
the dynamics of the entire system as the interaction 
between  an inner binary of point masses $m_{0}$ and $m_{1}$ and an external
binary of masses $M_{b}$ and $m_{2}$. 
We  define the angular momenta $L_{1}$ and $L_{2}$ of the inner and
outer binary:
\begin{equation}
L_{1}=m_{0}m_{1}\left[\frac{Ga_{1}\left(1-e_{1}^{2}\right)}{M_{b}}\right]^{1/2}~,\end{equation}
and \begin{equation}
L_{2}=M_{b}~m_{2}\left[\frac{Ga_{2}\left(1-e_{2}^{2}\right)}{M_{b}+m_{2}}\right]^{1/2}~,\end{equation}
with $G$ the gravitational constant.
 The dimensionless angular momenta are
$\ell_1=L_1/L_{1,c}$ and $\ell_2=L_2/L_{2,c}$ with $L_{i,c}=L_i/\sqrt{1-e_i^2}$
the angular momentum of a circular orbit with the same semi-major axis, $a_i$.

If the changes in the orbital properties of the inner binary
caused by its gravitational interaction with the third body occur on a timescale longer than
both the inner binary orbital period, $P_1$, and the tertiary orbital period, $P_2$, it is convenient
to average the equations of motion over the rapidly varying mean
anomalies of the inner and outer orbits~\citep[e.g.,][]{2015MNRAS.449.4221H}.
The resulting double averaged Hamiltonian is $\mathcal{H}=kW$
with $k=3Gm_{0}m_{1}m_{2}a_{1}^2/[8M_{b}a_{2}^3(1-e_{2}^2)^{3/2}]$
and \citep{1976CeMec..13..471L,2002ApJ...576..894M} \begin{eqnarray}\label{Ham}
W(\omega_{1},e_{1}) & = & -2(1-e_{1}^2)+(1-e_{1}^2){\rm cos^2}I \\
&  & +5e_{1}^2{\rm sin}^2\omega_{1}({\rm cos^2}I-1).\nonumber \end{eqnarray}

The quadrupole-level secular perturbation equations 
can be easily derived from the conserved Hamiltonian~(\ref{Ham}).
The resulting evolution equation of the inner binary orbital angular momentum due
to the torque from the tertiary  is\ \citep[e.g.,][]{1976CeMec..13..471L}: 
\begin{eqnarray} 
\frac{d\ell_{1}}{dt} &=&  -\frac{15 \pi}{4}\frac{m_2}{M_b}\frac{a_1^3}{a_2^3 (1-e_2^2)^{3/2}} \frac{ e_1^2}{P_1} 
\sin^2~I \sin 2 \omega_1 ~.\label{eq:eccen}
\end{eqnarray}
The characteristic timescale for the eccentricity oscillations is~\citep{1997Natur.386..254H}:
\begin{eqnarray} 
T_{\rm LK}\simeq P_{1}\left(\frac{M_b}{m_2} \right) \left(\frac{a_2}{a_1}\right)^3(1-e_2^2)^{3/2}~.
\end{eqnarray}

Using Equation~(\ref{eq:eccen}) and taking the relevant limit $e_1 \rightarrow 1$, 
we find that the  maximal 
change in the inner 
binary angular momentum due to the interaction with the third object and over the inner binary  orbital period is
\begin{equation}
\delta \ell_{\rm in}\approx 4\pi \frac{m_2}{M_{b}}\frac{a_1^3}{a_2^3\left(1-e_2^2\right)^{3/2}}\ ,
\end{equation}
and the  change in the 
binary angular momentum over one orbit of the tertiary  is
\begin{equation}
\delta \ell_{\rm out}\approx\delta \ell_{\rm in}  \frac{P_2}{P_1};
\end{equation} 
thus, the change experienced over the timescale associated with the period of the outer orbit
is naturally ${P_2}/{P_1}$ larger than the change experienced over the inner binary orbital period.

Equation\ (\ref{Ham}) describes the gravitational interaction of two weighted ellipses 
rather than  point masses on orbits and
is accurate as long as $\delta \ell_{\rm out} / \ell_1 \lesssim 1$, i.e., 
changes of the inner binary angular momentum 
occur on a timescale longer than the tertiary orbital period. 
If instead the change in the inner binary angular momentum 
over the inner and outer orbital periods is of order $\ell_1$,
then the system no longer meets the conditions required 
for the orbit averaged approximation, and more accurate direct
three-body integrations are needed.

Following the  discussion above,
in  order to address the reliability of the orbit averaged approach in describing the evolution
of a triple system it is useful to compare the change in angular momentum of the binary to 
the critical angular momentum $\ell_{\rm diss}\approx\sqrt{2D_{\rm diss}/a_1}$, where
$D_{\rm diss}$ is the scale at which other dynamical processes will dominate the evolution
of the inner binary. 
Since in this paper we 
focus on triples in which the inner binary can be driven to 
a strong close interaction or even a merger, we identify these processes  with 
energy loss due to GW radiation and tidal dissipation
in the case of BH and stellar binaries, respectively.

\begin{figure*}
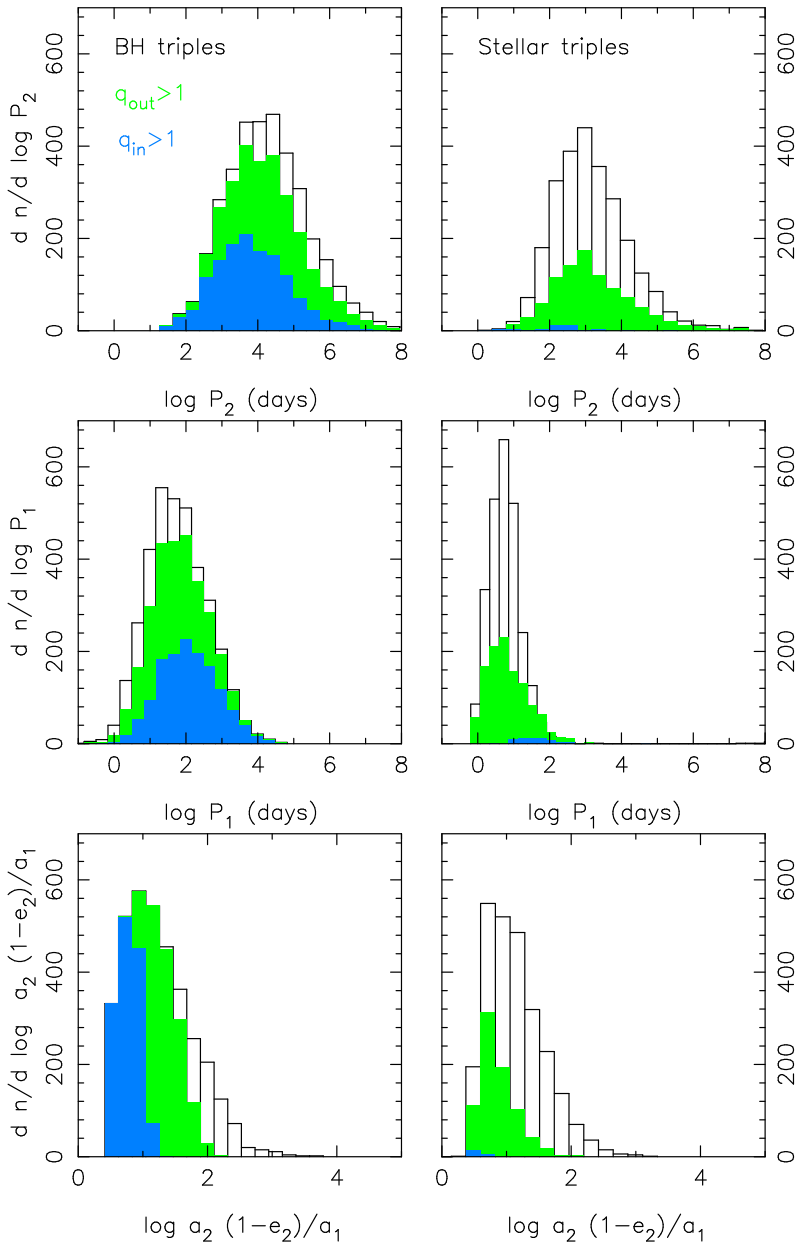

\centering
\includegraphics[width=2.15in,angle=270.]{Fig1a.eps}
\includegraphics[width=2.15in,angle=270.]{Fig1b.eps}
\includegraphics[width=2.15in,angle=270.]{Fig1c.eps}
\caption{Upper and middle panels 
show the distribution of inner and outer orbital periods of the BH and stellar triples 
formed in the Monte Carlo models of \citet{2015ApJ...800....9M}.
The bottom panel gives the distribution of the ratio between the
 outer binary periapsis and inner binary 
semi-major axis. Blue and green histograms indicate
 systems for which the assumptions on which the orbit averaged approximation is based  on break down before 
 GR or tides can affect their evolution.
Green histograms are systems for which  $q_{\rm out} \equiv {\delta \ell_{\rm out}}/{\ell_{\rm diss}}\gtrsim 1$;
blue histograms are systems for which  $q_{\rm in} \equiv {\delta \ell_{\rm in}}/{\ell_{\rm diss}}\gtrsim 1$.
For many triples formed through dynamical
interactions in star clusters, the secular approximation  breaks down
well above their typical dissipation scale, 
the LK equations of motion (even at the octupole order level) cannot 
 accurately trace the evolution of these systems.}\label{period-pdf}
\end{figure*} 

\begin{figure}
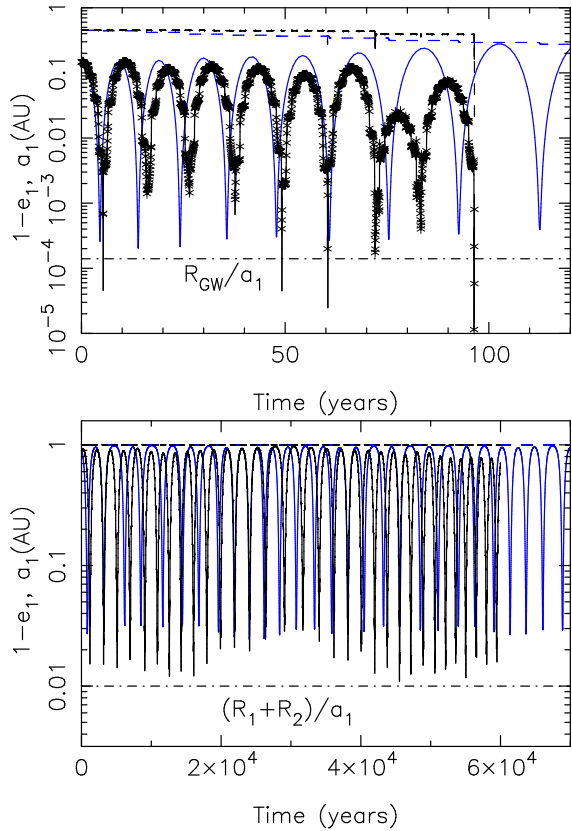

\centering
\includegraphics[width=2.15in,angle=270.]{Fig2a.eps} \\
\includegraphics[width=2.15in,angle=270.]{Fig2b.eps}
\caption{Dynamical evolution
of a BH triple (upper panel) and a stellar triple (lower panel). 
Black lines show the results of three-body integrations, 
blue lines show the results of the octupole order
orbit averaged equations of motion. 
Solid lines give the value of $(1-e_1)$,
 dashed lines give the inner binary semi-major axis.
In the upper panel the initial conditions  for the BH triple system were:
$m_0=14.9~M_{\odot}$, $m_1=13.8~M_{\odot}$, $m_2=14.2~M_{\odot}$
$a_1=0.45~{\rm AU}$, $a_2=3.9~{\rm AU}$,
$e_1=0.85$, $e_2=0.51$, $I=92.5^{\circ}$, 
$\omega_1=\omega_2=\pi/2$.
In the lower panel the initial conditions  for the stellar triple were:
$m_0=0.7~M_{\odot}$, $m_1=0.5~M_{\odot}$, $m_2=2~M_{\odot}$
$a_1=1~{\rm AU}$, $a_2=25~{\rm AU}$,
$I=101.8^{\circ}$, $e_1=0.1$, $e_2=0.7$, 
$\omega_1=214.6^{\circ}$, $\omega_2=0$.
In the upper panel the asterisks denote the minimum separation attained 
by the two BHs every orbital period. At 
$(1-e_1)\lesssim 10^{-3}$ this system 
satisfies the condition $q_{\rm in}\gtrsim1$, so that
the jump in periapsis over one inner orbit is of order the periapsis distance itself.
In the lower panel at $(1-e_1)\lesssim 0.04$ 
the triple  satisfies the condition $q_{\rm out}\gtrsim1$, 
so that the averaging procedure over the outer orbit is no longer justified.
In both examples the inner binary  reaches  higher eccentricities in
the three-body simulation than in the secular integration. This
allows the two BHs in the upper panel to merge through GW energy loss, and
the two stars in the lower panel to approach each other close enough 
that mass transfer would commence.
The dot-dashed lines indicate the critical value of $(1-e_1)$ below which 
GW radiation will dominate (upper panel),  and below which
the two stars will collide (lower panel).
These two examples clearly illustrate 
the limitation of the standard secular approach
even when including octupole order terms to the equations of motion.}\label{examples}
\end{figure}

The condition that the 
inner binary  undergoes rapid eccentricity oscillations 
over the \emph{outer} orbital period can be written as 
$q_{\rm out} \equiv {\delta \ell_{\rm out}}/{\ell_{\rm diss}}\gtrsim 1$;
this requirement translates into a condition on the outer perturber 
separation,
\begin{equation}\label{bd1}
\frac{a_2}{a_1}\lesssim  \frac{3}{{1-e_2^2}} \left(\frac{m_2}{M_b}\right)^{2/3}
\left(\frac{M_b}{M_b+m_2}\right)^{1/3}	
\left(\frac{a_1}{D_{\rm diss}}\right)^{1/3}~.
\end{equation}
We call this regime ``moderately'' non-secular.
In this regime systems can experience 
eccentricity oscillations over the outer binary period
and can be driven  to higher eccentricities than secular theory
would otherwise predict, which in the case of compact object binaries 
could result in substantially reduced merger times~\citep{2014MNRAS.439.1079A}.

If $q_{\rm in} \equiv {\delta \ell_{\rm in}}/{\ell_{\rm diss}}\gtrsim 1$,  
the inner binary will  undergo eccentricity oscillations on
the timescale of the \emph{inner} orbital period;
this can be achieved if the tertiary  comes closer to the inner binary than the separation
 \begin{equation}\label{bd2}
\frac{a_2}{a_1}\lesssim \frac{2.5}{1-e_2} 
\left( \frac{m_2}{M_b} \right)^{1/3}\left( \frac{a_1}{D_{\rm diss}}\right)^{1/6}\ .
\end{equation}
We call the regime defined by this equation ``strongly'' non-secular\ (see also \citet{2012ApJ...757...27A}
 and equation 7 in \citet{2012arXiv1211.4584K}).
In this situation the outer perturber is close enough to significantly
change the angular momentum  of the eccentric binary
at the last apoapsis passage leading to a
jump in angular momentum of order $\ell_1$. 
The angular momentum  of an eccentric inner binary can  
 jump from several times $\ell_{\rm diss}$  (far
enough to avoid GR or tidal effects)  to an arbitrarily small value.
Equation~(\ref{bd2}) defines
 the region of parameter space within which a compact object binary might 
 be expected to enter 
the aLIGO 
frequency band with a finite eccentricity~\citep{2012ApJ...757...27A,2013PhRvL.111f1106S,2014ApJ...781...45A}
or, in the case of inner stellar components, the binary members might experience a 
close interaction leading to a stellar collision~\citep{2012arXiv1211.4584K,2015ApJ...799..118P}.

Here we quantify the role of non-secular effects for triple systems forming in stellar clusters
by analyzing the properties of all dynamically stable triples that were produced in the GC
Monte Carlo models of \citet{2015ApJ...800....9M} (see Section\ \ref{MC-models}).  
The Monte Carlo models give the masses, stellar radii, semi-major axes, eccentricities and stellar type of all triples
formed during the cluster dynamical evolution;
these properties  were used to make Figure\ \ref{period-pdf} which shows
the period distributions as well as the distribution of the ratio $a_2(1-e_2)/a_1$
of all  BH triples (left panels) and stellar triples (right panels) that formed in these models 
\footnote{Hereafter we define as a stellar triple  any stable triple in which 
both components of the inner binary are MS stars; BH triples 
are identified with triples in which the two inner components are
both stellar BHs.}.

in Figure\ \ref{period-pdf} we have distinguished
 triples which satisfied the criterion~(\ref{bd1}) (green histograms) and  those 
meeting the more stringent condition equation~(\ref{bd2}) (blue histograms)
from the remaining systems which satisfied neither conditions.
In the case of stellar triples we 
evaluated equation~(\ref{bd1}) and (\ref{bd2})
setting $D_{\rm diss}=2(R_0+R_1)$ with $R_0$ and $R_1$
the physical radii of the inner binary components~\citep{2012arXiv1211.4584K}, while for 
BH triples  we adopted a conservative dissipation scale 
of $D_{\rm diss}=10^9\ {\rm cm}$~\citep{2014ApJ...781...45A}. 

Figure\ \ref{period-pdf} shows that for the majority of BH triples formed
non-secular dynamical effects are expected to become important
 before GR terms can significantly affect the evolution of the inner binary. 
We conclude that the standard  LK secular theory will fail in describing the evolution of these systems
due to the breakdown of the double averaging approximation.

Only a few stellar triples in our sample satisfy the condition~(\ref{bd2}),
mainly because of  their  large value of $D_{\rm diss}$.
However a large fraction of stellar triples 
still met the condition equation~(\ref{bd1}), meaning that the averaging procedure 
over the outer orbit is not justified for many of these systems.

In what follows we study the dynamical evolution of
 triples that form
through dynamical interactions in the dense stellar environment of GCs.
 We use a high accuracy three-body integrator to
study how the inner binary in the triples might evolve to become 
a possible source of GW radiation  for aLIGO detectors,
or, in the case of stellar binaries, how the evolution might lead to a 
mass transfer event or even a stellar merger. We begin in the next section by
describing the initial conditions and numerical methods adopted in our study.

\section{Methods and initial conditions} \label{meth+ICs}

\subsection{Numerical Methods and example cases}
The direct three-body integrations presented below were performed using the
ARCHAIN integrator~\citep{2008AJ....135.2398M}.
ARCHAIN employs an
algorithmically regularized chain structure and the time-transformed
leapfrog scheme to accurately integrate the motion of  
arbitrarily tight binaries with arbitrarily large mass-ratio.
The code also  includes post-Newtonian~(PN) non-dissipative 1PN, 2PN
and dissipative 2.5PN  corrections to all pair-forces. 
We refer the reader to  \citet{2008AJ....135.2398M} for  
a more detailed description of  ARCHAIN. 
We also used the octupole level secular equations of motion given in 
\citet{2002ApJ...578..775B}  to evolve the systems, which allowed us to directly compare the
predictions of the direct integrations  to the orbit averaged results.

In the case of triples containing an inner stellar binary (stellar triples),
to the PN terms we also added terms that account for 
dissipative tides as well as
apsidal precession 
induced by tidal bulges.
In order to do so we modified ARCHAIN including terms to the equations of motion
which represent tidal friction and quadrupolar distortion. 
Velocity-dependent forces were included via the
generalized mid-point method described in \citet{2006MNRAS.372..219M};
this method allows to time-symmetrize the algorithmic regularization
leapfrog even when the forces depend on velocities, permitting
the efficient use of extrapolation methods.
The tidal perturbation force has the form
\begin{equation} \label{tides-f}
\mathbf{F}=-G\frac{m_0m_1}{r^2} \left\{3 \frac{m_0}{m_1}
\left(\frac{R_1}{r}\right)^5 k \left( 1+3\frac{\dot{r}}{r}\tau \right)\hat{r} \right\}\ ,
\end{equation}
where $\tau$ (here set to  a fixed value of $1$\ sec) is the time lag,
and $k$  (set to $0.03$) is the apsidal motion constant.

Equation~(\ref{tides-f}) differs from the classical treatment of near-equilibrium tides
to the extent that we have
suppressed terms that are due to the difference in rotation rate
between the binary and each component. 
The orbital elements evolution equations 
that correspond to Equation~(\ref{tides-f}) are given by Equations~(9)
and (10) of \citet{1981A&A....99..126H}; these standard equations were implemented
 in the secular integrations to account for dissipative as well as non-dissipative tides.

To illustrate the importance of non-secular effects for the  dynamics
of the systems under consideration,
we show two examples in Figure\ \ref{examples}.
We consider a BH triple that is in
the strongly non-secular regime defined by Equation~(\ref{bd2}),
as well as a stellar triple  that 
is in the moderately non-secular regime defined by Equation~(\ref{bd1}).
The evolution of the two systems
was computed using both  the secular equations of motion
and ARCHAIN.

The examples of Figure\ \ref{examples} demonstrate how the double-orbit-averaging procedure 
can lead to misleading results when applied to the dynamics of systems
that evolve to attain very high orbital eccentricities.
In the upper panel the three-body integration predicts
that the inner binary will merge within a few LK oscillations
whereas the secular calculation predicts that the system does not merge 
in the considered time interval. More importantly, 
in the direct three-body integration the binary was found to have
a substantial eccentricity of  $e_1\approx 5\times 10^{-2}$ at the moment its 
peak gravitational wave frequency (Equation\ [\ref{gw_f}] below) reached  $10\ $Hz.
In the lower panel the stellar binary also attains higher eccentricities when integrated with 
ARCHAIN, which might induce a mass transfer event or reduce the circularization time
due to tidal friction when compared to the secular integrations.

\subsection{Globular cluster  models and initial conditions} \label{MC-models}
The initial conditions for the triple systems used in this analysis were taken from the Monte
Carlo simulations of \citet{2015ApJ...800....9M}. This study presents 42  Monte Carlo 
dynamical simulations of massive star clusters containing large populations of stellar BHs. 
The Monte Carlo technique simulates two-body relaxation in the Fokker-Planck approximation through 
representative pair-wise scattering interactions\ \citep{1971Ap&SS..14..151H,2010ApJ...719..915C}.

Our implementation in the Cluster Monte Carlo (CMC) code\ \citep[][and references therein]{2013ApJS..204...15P}
 includes direct integration of strong three- and four-body (binary--single and binary--binary) encounters, direct physical collisions,
 single and interacting binary stellar evolution, and tidal interactions with the Galaxy. 
Primordial triples are not included in these models.
However, during strong binary--binary interactions, 
it is possible to form stable hierarchical 
triple systems \citep{1995ApJ...438L..33R}.
Limitations in CMC currently require that these triples be broken artificially
 at the end of the timestep. Nonetheless, whenever a stable triple is formed, its properties are logged, 
including the masses, radii, and star types for the stellar components, and the semi-major axes 
and eccentricities for the inner and outer orbits.  
Since we lack the information regarding the mutual orientation of the two orbits,
we sample  $\omega_{1(2)}$ and $\cos(I)$ randomly from a uniform distribution with
$65^\circ \leq I \leq 115^\circ$.  The initial orbital phases in the three-body integrations
were also randomly distributed. 

We extracted the properties for every stable
 triple system formed through dynamical interactions 
over 12 Gyr across all of the 42 models presented in Morscher et al. 2015, 
and from five additional models provided by the same authors (Morscher, private communication).
The main properties of the 47 cluster models are reported in \citet{2015arXiv150500792R};
these span a large range of masses (from $2\times 10^5\ M_{\odot}$ to $1.6\times 10^6\ M_{\odot}$),
of virial radii (from $0.5\ $pc to $4\ $pc), and metallicities (Z=0.0005, 0.0001 and 0.005).
We note that, because these triple systems are destroyed in the Monte Carlo simulation, 
it is possible for the components of these triples to subsequently form new triple systems,
 when in reality they could in principle survive for a significant period of time. 

In the high-density stellar environment of GCs, triples may 
be perturbed through encounters with other passing stars on timescales 
that can be shorter than the relevant LK timescale.
Such encounters will alter the orbital properties of the triple significantly, or even disrupt it.
To account for this we set the final
integration time in our simulations equal to the encounter timescale
for collisions with other stars~\citep{1987gady.book.....B,2008MNRAS.386..553I},
\begin{eqnarray}\label{coll}
T_{\rm enc}&\approx& 8.5\times 10^{12}\ {\rm yr} P_2^{-4/3}M_{\rm tri}^{-2/3}\sigma_{10}^{-1} n_5^{-1} \nonumber \\
&&\times \left[1+913\frac{M_{\rm tri}+\langle M \rangle}{2P_2^{2/3}M_{\rm tri}^{1/3}\sigma_{10}^2}\right]^{-1}
\end{eqnarray}
where $\sigma_{10}$ is the central velocity dispersion of the cluster
in units of $10\ {\rm km~s^{-1}}$, $M_{\rm tri}=M_b+m_2$ is the mass of the triple, 
$\langle M\rangle$ is the mass of an average star in the cluster, and $n_5$ is  the local
number density of stars in units of $10^5\ {\rm pc^{-3}}$.
The  values of $n$, $\langle M\rangle$ and $\sigma$ in equation~(\ref{coll}) were obtained 
directly from the Monte Carlo models and correspond to  
to their values in the cluster core at the moment the triple system
was formed. 
We then integrate each triple until a time $T_{\rm enc}$ was reached,
or until the inner binary had merged.

 In total, the Monte Carlo models contain 8864 triples. Of these
5238 are BH triples; 3626 have at least one non-BH component in the inner binary, and
of these, in the inner binary, 2940 have two MS stars,
260 have a BH plus another MS star, 
218 have a white dwarf plus another MS star, 4 have a neutron star (NS) plus a MS star,
163 have a MS star plus an evolved star (red giant or later type), 
2 have a BH plus an evolved star. 
The remaining systems include
32 triples with an inner white dwarf binary,
5 triples with an inner BH plus white dwarf binary,
and 2 triples with an inner BH plus NS binary.
No triple with an inner NS binary is seen in any of these Monte Carlo models.
Of the 5238 BH triples, 3852 satisfy the condition $T_{\rm LK}\geq T_{\rm enc}$,
whereas  2519 of the 2940 stellar triples with an inner MS binary 
satisfied this condition.

\begin{figure}
\centering
\includegraphics[width=2.6in,angle=270.]{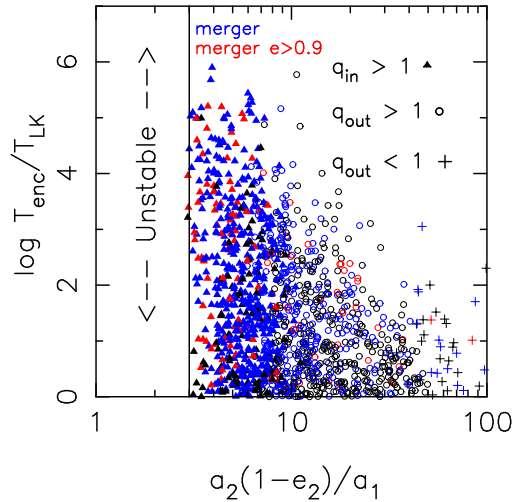}  
\caption{Ratio of the triple survival time in the GC core to the
LK timescale as a function of the ratio $a_2(1-e_2)/a_1$.
Triangle symbols represent systems that satisfy the condition Equation\ (\ref{bd2}),
open circles are systems which satisfy the condition Equation\ (\ref{bd1}) 
but not the condition Equation\ (\ref{bd2}) and
plus symbols are systems which satisfy neither of these two conditions.
We evolved each of these systems ten times selecting randomly the initial 
inclination as detailed in the text.
Blue colored points are systems in which in at least of the 10 realizations
we found one merger 
within $T_{\rm enc}$. Red points are systems in which 
for at least one of the ten realizations
the inner binary 
eccentricity was larger than $0.9$ at the moment its peak GW frequency
became larger than  $10$Hz. As predicted,
most eccentric mergers occur in triples for which the
 orbit averaging over the inner binary breaks down.
}\label{Fig3}
\end{figure}

\begin{figure*}
\centering
\includegraphics[width=2.7in,angle=270.]{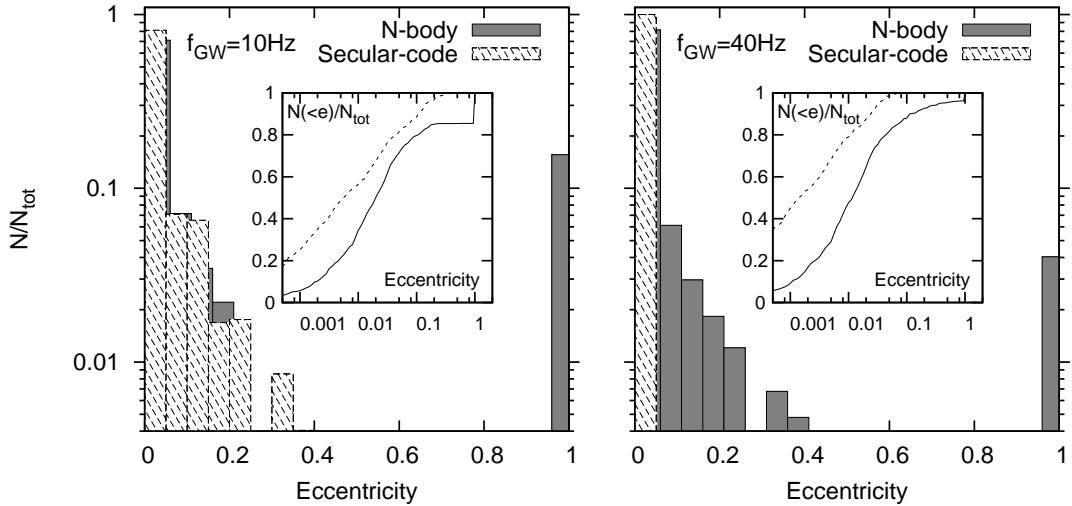}  
\caption{ Eccentricity distribution of merging BH binaries at the moment 
they first enter the $10$Hz (left panel) and $40$Hz (right panel) frequency bands.
While the \emph{octupole} level secular equation of motions 
predict that only a few percent of systems will have  finite eccentricity 
as they enter the aLIGO band, accurate $N-$body integrations
show that $\sim 20\%$\  ($\sim 10\%$)  of BH mergers in GC
will have an eccentricity larger than $0.1$ at $10$Hz\ ($40 \rm Hz$) frequency. 
About $10\%$\ ($\sim 5\%$) of all mergers will have extremely high eccentricities, i.e. $1-e\lesssim 10^{-4}$,
at $10$Hz\ ($40 \rm Hz$) frequency.
Note that the stippled regions
are in front in both panels, which means that 
the lack of stippled regions at high eccentricities is because there are none,
 rather than because they are hidden behind the solid hystograms.
}\label{Fig4}
\end{figure*}

\section{Results: black hole triples}\label{res-bhs}
For each of the 3852 BH triples extracted from the Monte Carlo models 
we made ten realizations, each adopting a random orientation of the two orbits and
random orbital phases.
Of the total 38520 BH triples that we evolved with ARCHAIN,
2080 resulted in a merger of the two
inner BHs. The total number of BH mergers obtained when using the orbit averaged 
equations of motion was 1796 instead. 
We found that the merger time of such binaries after the triple was formed was typically very short
 and always shorter than $\lesssim 10^5\ $yr. Thus, the BH mergers discussed in what follows 
will  occur inside the GCs even when the recoil velocity at formation can be large enough to eject the triple from the cluster core.
In this section we discuss the properties of the merging compact binaries
including their eccentricity, mass, merger time distribution, and event rates. 

Eccentric binaries emit a GW signal with a broad spectrum of frequencies.
the peak gravitational wave frequency  associated with 
the harmonic which leads to the maximal emission of GW radiation
can be estimated as~\citep{2003ApJ...598..419W} 
\begin{equation}\label{gw_f}
f_{{\rm GW}}=\frac{\sqrt{GM_{b}}}{\pi}\frac{~~(1+e_{1})^{1.1954}}{\left[a_{1}(1-e_{1}^{2})\right]^{1.5}}~.
\end{equation}
We  approximately follow the time evolution of 
the GW frequency of the inner BH binaries  
through Equation~(\ref{gw_f}). This allowed us to estimate the eccentricity of the BH binary
when its GW frequency is $\gtrsim 10~$Hz, i.e., 
when it would be large enough to be into the aLIGO frequency band\ \citep{2010CQGra..27q3001A}.

Figure~\ref{Fig3} gives the ratio of the triple survival timescale, 
$T_{\rm enc}$, to the LK timescale plotted against 
the value of $a_2(1-e_2)/a_1$. 
From this plot we see that non-secular dynamical effects are expected to become important to
the evolution of most BH triples in our models, possibly leading to the formation of eccentric GW sources.
Also, most BH mergers are found to occur in moderately hierarchical triples with $a_2(1-e_2)/a_1\lesssim 10$.
This is expected for at least two reasons: (1)
the closer the outer body to the inner binary the less significant  is
the quenching of the LK cycles due to relativistic 
precession of the inner binary~\citep[e.g.,][]{2002ApJ...578..775B}, so that typically
the smaller  the ratio $a_2(1-e_2)/a_1$ the larger the maximum eccentricity attained by the inner binary and,
consequently, the shorter its merger time; (2) triples with small $a_2(1-e_2)/a_1$ ratio, have also
larger $T_{\rm enc}/T_{\rm LK}$ ratio, which naturally leads to a higher chance for a merger before the triple is
disrupted by gravitational encounters with other stars. 
In Figure~\ref{Fig3} we also identify those systems for which at least one of the 10 random realizations 
led to a merger (blue points) and those for which at least one realization 
produced a BH  binary merger with an eccentricity larger than 0.9 at $\gtrsim 10\ $Hz frequency (red symbols).
As  predicted on the basis  of our discussion in Section~\ref{bd-oaa}, most eccentric mergers occur 
at $a_2(1-e_2)/a_1\lesssim 10$, near the boundary for instability.

\begin{figure}
\centering
\includegraphics[width=2.3in,angle=270.]{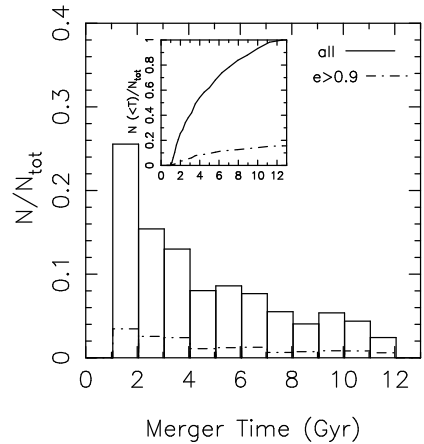}  
\caption{Merger time distribution of BH binaries due to the LK
mechanism in GCs. Continue lines give the merger time distribution
of all BH binary mergers, dashed line is for BH binaries
that have $e>0.9$ when $f_{{\rm GW}}\geq 10\ $Hz.
Insert panel gives the corresponding cumulative distributions.
}\label{Fig5}
\end{figure}

\begin{figure}
\centering
\includegraphics[width=2.3in,angle=270.]{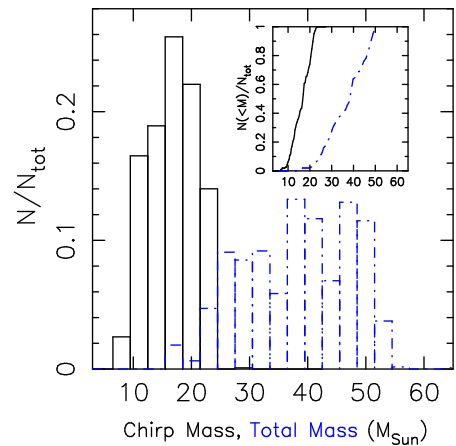}  
\caption{Continue black lines show the chirp mass distribution of
BH binary mergers due to the LK mechanism in GCs.
Dot-dashed blue lines give the distribution of total mass for the merging binaries.
Insert panel gives the corresponding cumulative distributions.
}\label{Fig6}
\end{figure}

Figure~\ref{Fig4} shows the eccentricity distribution of all BH binary mergers in our simulations,
when the associated $f_{{\rm GW}}$ \emph{first} enters the $10~$Hz 
and $40~$Hz frequency bands. Figure~\ref{Fig5} and \ref{Fig6} show the corresponding merger time and mass distribution
of the merging binaries. The distributions showed in Figures \ref{Fig4}, \ref{Fig5} and \ref{Fig6} 
were obtained by weighing the number of mergers for each cluster model by the likelihood of that
model to represent a typical  GC in the Milky Way\ (MW). 
More in detail, the weights are
obtained by creating a kernel density  estimate ($W_{\rm MW}$) of the MW GCs on 
the fundamental plane (which we take to be mass and ratio of the half to core radius), 
then estimating the weight for each model using the kernel density estimate at the position
of that model on the fundamental plane. In this way, cluster models that are more likely to be drawn from the
same distribution of MW GCs are more heavily 
weighted.
The weight for a model of total mass $M_{\rm GC}$, core radius $R_c$ and half mass radius $R_h$ after $12\ $Gyr of evolution
is computed as:
\begin{equation}\label{wgts}
W(M,R_c/R_h)={W_{\rm MW}(M,R_c/R_h)\over W_{\rm Models}(M,R_c/R_h)}~,
\end{equation}
where we have divided by $W_{\text{Models}}$, the kernel density estimate of the models themselves.  
This serves to normalize the distribution, so that regions of parameter space that are over-sampled by the 
models are given lower weights
~\citep[see][for more details]{2015arXiv150500792R}. 
The  fraction of mergers with a given property (e.g., total mass, eccentricity) is then simply
$f=\sum_i N_i W_i/\sum_i W_i$, with $N_i$ the number of mergers 
occurring in the $i_{\rm th}$ cluster model.

\citet{2013PhRvD..87l7501H} showed that 
for eccentricities less than $e_1\lesssim 0.1$ 
at $\approx 10\ $Hz, circular templates 
will be effective at recovering the GW signal of eccentric sources.
Figure~\ref{Fig4}  shows that approximately $20\%$ of all BH mergers 
in our three-body integrations
had an eccentricity $e>0.1$ at $\gtrsim 10~$Hz frequency. This percentage drops to $\sim 10\%$ at  $40~$Hz frequency.
The difference with the results of the secular equations of motion is evident in these plots.
The secular integrations clearly underestimate the number of eccentric mergers producing just a few
percent of inspirals with $e\gtrsim 0.1$ at $10~$Hz frequency. 
The direct three-body integrations also 
produce a significant population ($\sim 15\%$ of the total) of highly eccentric sources in the aLIGO
frequency window, which are fully missed when evolving the triples with the secular equations of motion.
These sources  will start to inspiral due to GW radiation energy loss within the aLIGO band with an 
extremely high eccentricity, $1-e_1\lesssim 10^{-4}$, and a characteristic
GW frequency.  In fact,  the angular momentum of BH binaries that evolve into the strongly non secular regime
can be arbitrarily small at the moment  energy loss due to GW radiation  
starts to become dominant. For this reason the GW frequency at which the binary enters the aLIGO band,
which depends on the binary angular momentum  via Equation~(\ref{gw_f}),
could be significantly larger than $10\ \rm Hz$.
 \citep[see the example of Figure 3 in][]{2014ApJ...781...45A}.
We note that the detectability of such sources also depends on how much energy is
radiated around $100\ \rm Hz$ where aLIGO is more sensitive. So it is important to see
at what frequency these sources first enter the aLIGO sensitivity window.
Based on our simulations we found that all eccentric BH binaries will enter the aLIGO band at $\lesssim 300\ \rm Hz$
and $80\%$ of them had $f_{\rm GW}<50\ \rm Hz$ when they first enter the aLIGO frequency band\ \citep[see also Figure 7 in][]{2014ApJ...781...45A}. 
We conclude that most eccentric sources could be detected by aLIGO, provided that an efficient search strategy is used.

Figure\ \ref{Fig5} displays the average merger time distribution of BH binaries
induced by the LK mechanism in our models. 
The figure shows a clear peak around 1Gyr of evolution
with a long tail extending to the present epoch.
The declining merger rate is mostly due to the decreasing total number of BHs in our models with time.
As the rate depends on the total number of BHs in the cluster core it drops with time as 
BHs are continuously ejected from the cluster through strong dynamical interactions. 

Figure\ \ref{Fig6} displays the chirp mass, $M_{\rm ch}=\left(m_0m_1\right)^{3/5}/\left(m_0+m_1\right)^{1/5}$, as 
well as the total mass distribution of the merging BH binaries. 
These distributions appear to be consistent with that
of dynamically formed binaries shown in Figure\ 2 of \citet{2015arXiv150500792R}.
Figure\ \ref{Fig6} shows a mass distribution peaked around $M_{\rm ch}\approx 17~M_{\odot}$,
while the highest chirp mass of BH binaries formed by pure stellar evolution over all our models was 
 $\approx 13~M_{\odot}$.  This latter value for the chirp mass is the maximum for
 primordial binaries in our GC simulations that evolved without significant dynamical
encounters~\citep[see also][]{2010MNRAS.407.1946D,2015arXiv150500792R}. 
The fact that the typical chirp mass of the merging BH binaries in Figure\ \ref{Fig6}  
is substantially larger than $\approx 13~M_{\odot}$
is a result of the naturally stronger segregation of  the
heaviest BHs as well as the preferential ejection of the lightest BHs from the clusters during dynamical interactions.
As the heavier BHs segregate more efficiently to the cluster center and the lighter BHs are ejected through
binary single interactions, the former will tend to dominate the cluster core, where
binary-binary interactions can lead to the formation of stable BH triples and to LK induced mergers.

\subsection{Black Hole Merger Detection Rate}\label{d-rates}
One of the advantages of using realistic cluster Monte Carlo models is that it
allows us to make direct predictions about the mass and merger time
distribution of the coalescing BH binaries, which is fundamental to make
reliable estimates for the event and detection rates.  As we have complete
information about the distribution of sources in time and chirp mass, we can
compare our ensemble of GC models to observations of MW and extragalactic
GCs and estimate the merger rate for a single aLIGO
detector.

To compute the rate of detectable sources per year from triple systems, we
follow a similar procedure to \cite{2015arXiv150500792R}.  
 The rate is expressed as 
the following double integral over source chirp mass and redshift:

\begin{equation}
R_d = \iint \mathcal{R}(\mathcal{M}_c,z) f_d(\mathcal{M}_c,z) \frac{dV_c}{dz} \frac{dt_s}{dt_0} d\mathcal{M}_c dz~.
\label{eqn:rate}
\end{equation}

where

\begin{itemize}
\item $\mathcal{R}(\mathcal{M}_c,z)$ is the rate of BH binary mergers with chirp
mass $\mathcal{M}_c$ at redshift $z$ due to the LK mechanism.  
\item $f_d(\mathcal{M}_c,z)$ is the fraction of sources with chirp mass $\mathcal{M}_c$ at redshift $z$ that are detectable by a single 
aLIGO detector.
\item $dV_c/dz$ is the comoving volume at a given redshift \citep{1999astro.ph..5116H}.  We
assume cosmological parameters of $\Omega_M = 0.309$, $\Omega_\Lambda = 0.691$,
and $h = 0.677$, from the combined Planck results \citep{2015arXiv150201589P}.
\item $dt_s/dt_0 = 1/(1+z)$ is the time dilation between a clock measuring the merger rate at the source versus a clock on Earth.
\end{itemize}

The rate is expressed as the product of the mean
number of triple sources per GC, the distribution of sources in
$\mathcal{M}_c-z$ space, and the spatial density of GCs per ${\rm Mpc^3}$ in the local
universe (0.77$\ {\rm Mpc}^{-3}$, from \cite{2015arXiv150500792R}, Supplemental
Materials).  Symbolically, this becomes $\mathcal{R}(\mathcal{M}_c,z) = \left<N\right> \times
P(\mathcal{M}_c,z)\times \rho_{GC}$.

\begin{table*}
\begin{center}
\caption{\scriptsize{aLIGO detection and merger rates of coalescing BH binaries from GCs.
(In-cluster mergers are defined here as all events occurring inside the GC Monte Carlo models that
are not due to the LK mechanism.
As described in the text, the detection rate for eccentric sources ($e_1[\gtrsim10\ {\rm Hz}]\gtrsim 0.1$) 
represents the number of inspirals that could be detected with
dedicated search strategies.)} } 
\begin{tabular}{l|ccc|ccc}
\hline 
Source  & \phantom  &Detection rates (yr$^{-1}$) & & &  Merger rates (${\rm GC^{-1}}$ Myr$^{-1})$  \\ 
\phantom  & low & realistic & high  & low & realistic & high   \\ 
 \hline
 Hierarchical triples             & 0.11& 0.47 & 1.93	   	&  $1.9\times 10^{-4}$   & $2.8\times 10^{-4}$  & $3.3\times 10^{-4}$  \\
 Eccentric inspirals    	 & 0.021& 0.20& 0.48	&  $3.4\times10^{-5}$  	& $5.7\times10^{-5}$ & $7.3\times 10^{-5}$  \\
In-cluster mergers              & 0.011& 0.062& 0.60	        &  $1.9\times 10^{-4}$     & $2.8\times 10^{-3}$  & $3.4\times 10^{-3}$    \\
Ejected binaries                  & 10& 30 &100  &  $0.015$         & 0.023  & 0.027    \\
\hline
\end{tabular}
\end{center}
\label{TAB1}
\end{table*} 

The components of $\mathcal{R}(\mathcal{M}_c,z)$ are computed as follows.
The mean number of sources per GC is found by performing a weighted linear
regression between the number of triple mergers a GC produces over its lifetime,
and the final mass of the GC at 12 Gyr.  
 The number of triple mergers for a given cluster
is obtained by multiplying the total number of mergers obtained from the three-body integrations
by a factor $\cos(65^{\circ})\times 0.1$; where the first term takes into account
the fact that we have only considered inclinations $65^\circ \leq I \leq 115^\circ$ and 
the second term the fact that for each triple we have performed ten integrations with random orbital orientations 
and phases.
The weights are assigned using
$W_{\rm MW}(M,R_c/R_h)$, the numerator from Equation\ (\ref{wgts}).  
This linear regression is then multiplied by a universal GC luminosity
function \citep{2014ApJ...797..128H}, and integrated over mass (assuming a
mass-to-light ratio of 2) to yield a mean number of sources.  This average is then multiplied by the distribution of
sources, found by randomly drawing a number of mergers from each model according
to the weights in Equation\ (\ref{wgts}), then generating a kernel density 
estimate
of these sources in 
chirp mass/redshift space. Note that unlike \cite{2015arXiv150500792R}, we do not
separately consider the populations of low-metallicity and high-metallicity GCs.

Finally, we model the efficiency of an aLIGO detector by calculating
the fraction of sources at chirp mass $\mathcal{M}_c$ and redshift $z$ that aLIGO
could detect, marginalized over all possible sky locations and binary
orientations.  We use IMRPhenomC waveforms \citep{2010PhRvD..82f4016S} and the
projected zero-detuning, high-power aLIGO noise curve \footnote{ \protect \href
{''https://dcc.ligo.org/cgi-bin/DocDB/ShowDocument?docid=2974''}{LIGO Document
T0900288-v3}}.   Note that this template family does \emph{not} consider the
eccentricity of sources, which could dramatically alter
the gravitational waveforms\ \citep{2013PhRvD..87l7501H}. Such simplification does not significantly 
alter the total detection rate estimates presented here; in fact, our simulations show that 
at $10\ {\rm Hz}$ frequency $\sim 80\%$ of coalescing BH binaries have an eccentricity 
$e_1\lesssim 0.1$, which is small enough that circular templates will be
efficient at recovering their GW signal \citep{2010PhRvD..81b4007B,2013PhRvD..87l7501H}.

As $P(\mathcal{M}_c,z)$ is dependent on the random draw of mergers, we
report the mean of $R_d$ over 100 separate draws.  We also provide optimistic
and pessimistic rate estimates to approximate the uncertainties associated with
our assumptions.  The
optimistic rates are computed by assuming the $+ 1\sigma$ value of the 100 draws
of $R_d$, the $+1\sigma$ uncertainty for the linear regression between GC mass and number of
mergers, a highly optimistic upper limit on the GC mass function ($2\times10^8
M_{\odot}$), and the upper-limit on $\rho_{GC}$ of $2.3\ {\rm Mpc}^{-3}$ from
\cite{2015arXiv150500792R}.  Conversely, the pessimistic rate estimate assumes the
corresponding $-1\sigma$ uncertainties, a pessimistic upper limit on the GC mass
function ($4\times10^6 M_{\odot}$), and the lower-limit of $\rho_{GC}$ of $0.32\
{\rm Mpc}^{-3}$.

For all LK driven events, we find that a single aLIGO detector could detect $\sim 0.5$ events
per year (pessimistically $\sim 0.1\ \rm yr^{-1}$, optimistically $\sim 2\ \rm yr^{-1}$).

We performed the same analysis described above for all eccentric sources produced in our 
three-body simulations,
i.e., we computed the detection rate solely for eccentric systems
 \emph{assuming perfect templates for detecting eccentric binaries}.
Thus, the results of this computation 
\emph{do not} represent the number of sources that aLIGO could detect with current searches.
Instead, they suggest an estimate of 
the number of sources that aLIGO could detect if
optimal matched-filtering searches for eccentric sources were to be used. 
For events with eccentricities greater than 0.1, this rate is
$\sim 0.2\ \rm yr^{-1}$ (pessimistically $\sim 0.02\ \rm yr^{-1}$, optimistically $\sim 0.5\ \rm yr^{-1}$), 
while for mergers with eccentricities greater than 0.9, the rate
becomes $\sim 0.2\ \rm {yr^{-1}}$ (pessimistically $\sim 0.01\ \rm yr^{-1}$, optimistically
$\sim 0.4\ \rm yr^{-1}$).  We note also that that  the corresponding waveforms
of most eccentric GW sources discussed here 
are characterized by a sequence of 
bursts of energy emitted near periapsis rather than a
continuous waveform. Because of this, matched filtering techniques could
be unpractical and a power stacking algorithm might be employed instead\ \citep{2013PhRvD..87d3004E}.
The disadvantage is that this will not give an optimal signal-to-noise ratio which might somewhat reduce 
the detection rates for eccentric inspirals we find.

Table\ 1 summarizes the main results of our rate computation giving
the predicted aLIGO detection rate of binary BH mergers from GCs.
This table also gives the corresponding compact binary coalescence rates per GC per Myr.
Table\ 1 gives the  rate of LK induced BH mergers\ (hierarchical triples),
and the rate of mergers due to the LK mechanism that have an
eccentricity larger than 0.1 at $10~$Hz frequency\ (eccentric inspirals).
From our Monte Carlo models we also computed the merger rate of
hard escapers that will merge in the galactic field within one Hubble time\ (ejected binaries)
and the rate of mergers occurring  \emph{inside} the clusters
that are not due to the LK mechanism\ (in-cluster mergers). 
These latter  include  mergers due to binary-single dynamical interactions, 
as well as those mergers due to binary stellar evolution and
single-single captures~ \citep[see][for details]{2015arXiv150500792R}. 

In agreement with previous calculations\ \citep[e.g.,][]{2006ApJ...637..937O,2010MNRAS.407.1946D} we find
that the total event rate is largely dominated by
outside cluster mergers. 
We stress again that the BH mergers due to the LK mechanism will occur inside the GC.
Thus, by comparing the rate estimates for triples to the rate of in-cluster mergers 
due to other processes in Table\ 1 
we can conclude that the detection rate of BH mergers occurring \emph{inside} the cluster
is dominated by  the rate of LK induced mergers.
This is somewhat surprising given that in-cluster mergers 
that do not occur in triples account for 
approximately $10\%$ of the total binary BH mergers over 12Gyr.
The low aLIGO event rate we find is a consequence of the fact that
the majority of these mergers occur early in the cluster's lifetime, and are relatively low-mass.
Since aLIGO is less sensitive to low-mass, high-redshift sources, the rate of detected sources drops.
We conclude that mergers induced by LK oscillations in hierarchical triples
are likely to be the main contributor to in-cluster gravitational wave sources for aLIGO detectors,
and represent $\sim 1\%$ of the anticipated total detection rate of BH binary mergers assembled in GCs.

\begin{figure}
\centering
\includegraphics[width=2.6in,angle=270.]{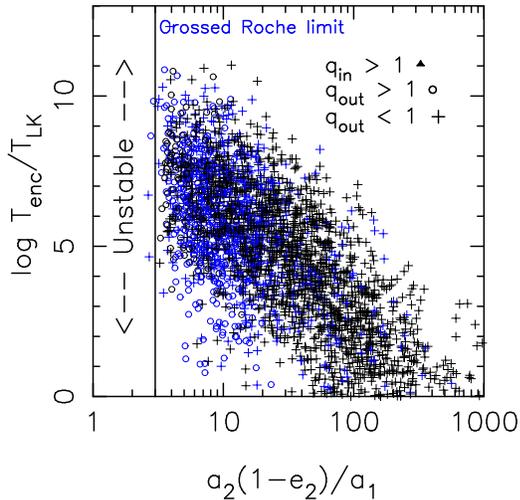}
\caption{Similar to Figure\ \ref{Fig3} but for stellar triples.
Blue symbols are systems in which one component of the inner binary crossed the
Roche limit during the evolution.
}\label{Fig7}
\end{figure}

\begin{figure*}
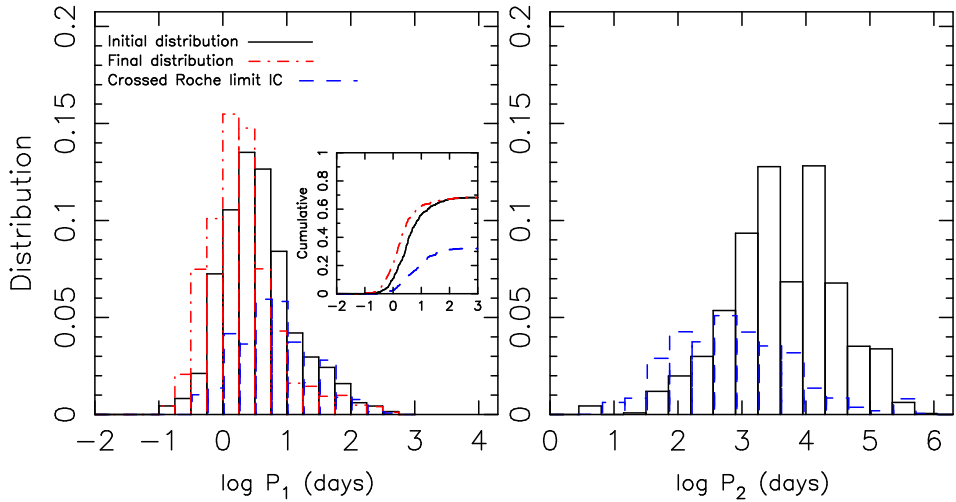

\centering
\includegraphics[width=2.6in,angle=270.]{Fig8a.eps}
 \includegraphics[width=2.6in,angle=270.]{Fig8b.eps}
\caption{Initial (black lines) and final (dot-dashed red lines) period distributions
for systems in which the inner binary components did not cross the Roche limit
during the evolution. Dashed blue lines show  the initial period distribution of
systems that crossed their Roche limit during the evolution.
The distributions are normalized to the total number of systems
that were evolved, such that the sum of the integrals of the distribution of not merging systems
plus the distribution of merging systems is unity.
As explained in the text these distributions should be interpreted as typical distributions for
MW-like GCs. 
}\label{Fig8}
\end{figure*}

\section{Results: stellar triples}\label{st-tr}
In total, our Monte Carlo models produced 2940 triples in which both
components of the inner binary were MS stars\ (i.e., stellar triples). 
Following the same procedure adopted above, we made ten realizations 
for each triple, taking a random  orientation between the two orbits 
and random orbital phases.
We then integrated the triples forward in time using ARCHAIN if
the condition\ (\ref{bd1}) (with $D_{\rm diss}=2[R_0+R_1]$) was satisfied;
the rest of the initial conditions were integrated using the octupole level secular equations
of motion based on the double average Hamiltonian~\citep[e.g.,][]{2014ApJ...793..137N,2014ApJ...794..122M}.
Each integration was terminated  either at $T_{\rm enc}$, or when  one of the binary components
had passed within  its Roche limit.

In many of our simulations the inner binary reaches very high eccentricities, 
implying a high probability that the stars will cross each other's Roche limit and transfer mass.
To account for the possibility of crossing the Roche limit, we 
check whether the condition 
$d_{ij}L_{R,ij}<R_i$ (or $a_1(1-e_1)L_{R,ij}<R_i$ in orbit average integrations) was met, 
with $d_{ij}$  the instantaneous distance between the 
two inner binary components and\ \citep{1983ApJ...268..368E}:
\begin{equation}\label{roche}
L_{R,ij}=0.49\frac{\left(m_i/m_j\right)^{2/3}}{0.6\left(m_i/m_j\right)^{2/3}+\ln \left(1+\left(m_i/m_j\right)^{1/3} \right)}\ .
\end{equation}
As often done in the literature, we stop the integration and assumed that the binary had merged
if one of the two binary components had passed within its Roche limit\ \citep[e.g.,][]{2014ApJ...793..137N}.
We found that in 10327 out of the overall 29400 triples, one of the inner binary components crossed its Roche limit.
After accounting for the fact that we have explored a limited range of orbital inclinations,
we find an overall probability  of $\sim 10\%$ that a stellar triple assembled dynamically in a GC will lead
to a mass transfer event and possibly to a stellar merger.

Figure~\ref{Fig7} gives the ratio of the triple survival timescale, 
to the LK timescale plotted as a function of
$a_2(1-e_2)/a_1$. Mass-transfer systems are indicated by blue colored symbols.
This plot shows that the survival time-scale for these systems can be many orders of magnitude 
larger than the corresponding LK time-scale.
 The  ratio $T_{\rm enc}/T_{\rm LK}$  for these systems is also typically larger than that for BH triples\ (compare to Figure\ \ref{Fig3}),
reflecting the different tertiary period distribution of the two populations shown in Figure\ \ref{period-pdf}
and the larger binary mass in the case of BH components.
In Figure\ \ref{Fig7} we identify those systems for which the orbit average approximation breaks down.
In agreement with our previous discussion, most of these systems correspond to compact configurations
in which the tertiary comes closer to the inner binary than a separation $a_2(1-e_2)/a_1\lesssim 100$.
We also identify those systems 
in which in at least  one of the ten random realizations one of the  inner binary components crossed 
its Roche limit.
Not surprisingly,  close outer orbits are more likely to result in a mass transfer event between the 
inner binary components.

\begin{figure}
\centering
\includegraphics[width=2.3in,angle=270.]{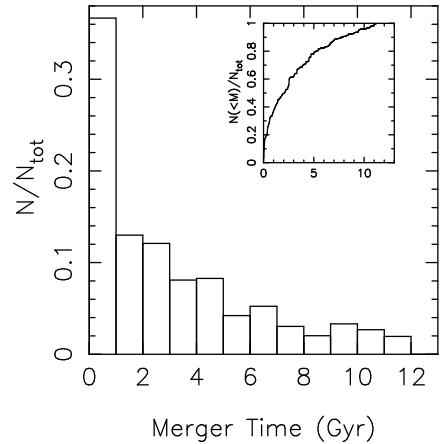}
\caption{ Merger time distribution of stellar binaries 
produced by the LK mechanism in a MW-like GC.
Insert panel shows the corresponding cumulative distribution.
The distribution has been normalized to the total number of merging systems.
}\label{Fig9}
\end{figure}

\begin{figure}
\centering
\includegraphics[width=2.3in,angle=270.]{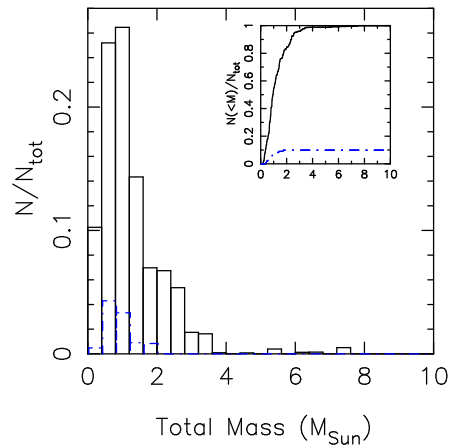}
\caption{Distribution of the total mass of binaries that crossed the Roche limit during the 
evolution. Assuming that these binaries will end up merging, such distributions can be interpreted as
those corresponding to stellar merger products induced by the LK mechanism in a MW-like GC.
The dashed-blue lines show the mass distribution of systems that merged in the last $4 $Gyr of the cluster evolution.
These distributions have been normalized to the total number of merging systems.
}\label{Fig10}
\end{figure}

In Figure~\ref{Fig8} we consider the distribution of orbital periods.
Figures~\ref{Fig9} and Figure~\ref{Fig10} display respectively the distribution of
merger times and mass of the merger remnant for systems that crossed their Roche limit.
These distributions were obtained using the weighing technique described in Section\ \ref{res-bhs},
i.e., weighing the number of systems for each cluster model by the likelihood of
that model to represent a MW-like GC.

In Figure~\ref{Fig8} we show the initial and final orbital period distributions 
of the systems that did not cross the Roche limit, as well as the
initial period distributions of the systems that crossed the Roche limit during the evolution.
As shown in this figure the peak of the  period distribution of the surviving binaries 
shifts from $\sim 3.2\ $days to  $\sim 1\ $day. This is a consequence of 
energy loss due to tidal dissipation that the binary experiences during the 
high eccentricity phases of LK cycles.
If during the evolution the  timescale associated with precession due to 
tidal and PN terms becomes shorter than the LK timescale, the eccentricity 
oscillations due to the interaction with the tertiary are suppressed.
In this case, after many LK cycles
tidal dissipation can shrink the binary orbital semi-major axis by 
a large factor. 
causing the binary to effectively decouple from the tertiary.

 Figure~\ref{Fig8} shows that  although merging systems are produced
at any value of $P_1$ and $P_2$, they
 are more likely to originate from systems
with the tightest outer orbits and widest inner binaries.
In fact, about half of the systems with
$P_2\lesssim 100\ $days or with $P_1\gtrsim 10\ $days  led to stellar mergers.
Generally in order for the inner binary to attain a quasi-radial orbit such that its components can
merge, its eccentricity has to be excited on a timescale much shorter than the typical 
extra precession timescale (such as tides and
relativistic precession). It can be easily shown that
the effectiveness of such  extra sources of precession decreases
with the ratio $a_2/a_1$\ \citep[e.g.,][]{2002ApJ...578..775B}, which is consistent with the fact that mergers
are more likely to occur for wide inner binaries and close outer orbits.

The  distribution of merger times for stellar binaries is shown in Figure~\ref{Fig9}.
Given that the merger time through LK cycles is generally very short, 
the number of merging systems at a given time is also a good measure of
the number of hierarchical triples that are formed dynamically in the cluster core at that time interval.
Figure~\ref{Fig9} shows that $\sim 40\%$ of all stellar mergers occur in the first Gyr of cluster evolution.
This is expected given that it is only during these early times that the stars dominate the cluster core where
 stellar binary-binary encounters can take place at a high rate and lead to the assembly of
stellar triples. After $\sim 1\ $Gyr of evolution, the BHs have already decayed to the center through dynamical friction 
becoming the dominant population in the cluster core. As the BHs become dominant, interactions 
involving  stellar binaries become less frequent, naturally reducing  the formation rate of triples with stellar components
and favoring the formation of BH triple systems instead. 

Finally, Figure~\ref{Fig10} displays the total mass distribution of all binaries that 
underwent Roche-lobe overflow during their
evolution. This plot shows that the merging systems have masses up to $\approx 8\ M_{\odot}$
 with a peak at about $1\ M_{\odot}$. 
The most massive mergers, including a few at $\gtrsim 4\ M_{\odot}$,  were found to occur 
only very early in the first few Gyrs.
As illustrated by the dashed-blue lines in Figure~\ref{Fig10},
no merger with mass larger than 
$\sim 2\ M_{\odot}$ was found during the last $4\ $Gyr of cluster evolution.
These late mergers have a typical mass of $0.5-1\ M_\odot$, as expected being about twice the mass of an
average MS star of an old single stellar population.

\subsection{Formation of blue stragglers}
BSSs are main sequence stars that appear to be hotter and 
more luminous than the turn-off point for their parent cluster 
population. As such they appear  to be  younger than the rest of the stellar population \citep[e.g.,][]{1953AJ.....58...61S}. 

BSSs are created when a MS star is replenished with new supply of Hydrogen in its core. Hence, 
they can remain in the MS longer than expected of an undisturbed star. 
This can happen through mass transfer in a binary system or via merger/collision with at least 
one MS star  
\citep[see][for a recent review]{2015ebss.book..251P}.
\citet{2009ApJ...697.1048P} suggested that triples might be a natural progenitor for
BSSs, as the high eccentricities attained during LK  cycles can lead to tidal interaction,
mass loss or even a merger of the inner binary components. Although dynamically formed triples
are expected to form quite efficiently in GCs, their long term dynamical evolution in GCs has not been yet investigated. 
Hence, until now their role in BSS formation remains largely unconstrained.
In what follows we use  our Monte Carlo models coupled with the results of the three-body integrations to 
determine the relative contribution of triples to  BSSs in GCs.

Besides BSSs possibly produced by the LK mechanism,
we also computed the number  of BSSs due to binary-mediated interactions, i.e., mass transfer and physical collisions.
We computed the number of BSSs at the observation time of $\sim12\,\rm{Gyr}$ in each model in the following way. 
We first estimate the MS turn-off mass ($m_{\rm{TO}}$) given the metallicity and age of the model cluster in question. 
From the snapshot containing all stars we then extract those that are still on their MS and has mass 
$\geq 1.05\times m_{\rm{TO}}$. In theory, any MS star with mass above $m_{\rm{TO}}$ are BSSs, however, 
observationally, it is hard to distinguish BSSs from the MS if their masses are not high enough compared to $m_{\rm{TO}}$. 
This simple mass-based 
prescription results in very good agreement between observationally extracted BSS numbers, their stellar properties, 
and known correlations between BSS and observable properties of the cluster 
\citep[e.g.,][for more details]{2011MNRAS.416.1410L,2013ApJ...777..106C,2013ApJ...777..105S}. 

\begin{figure*}
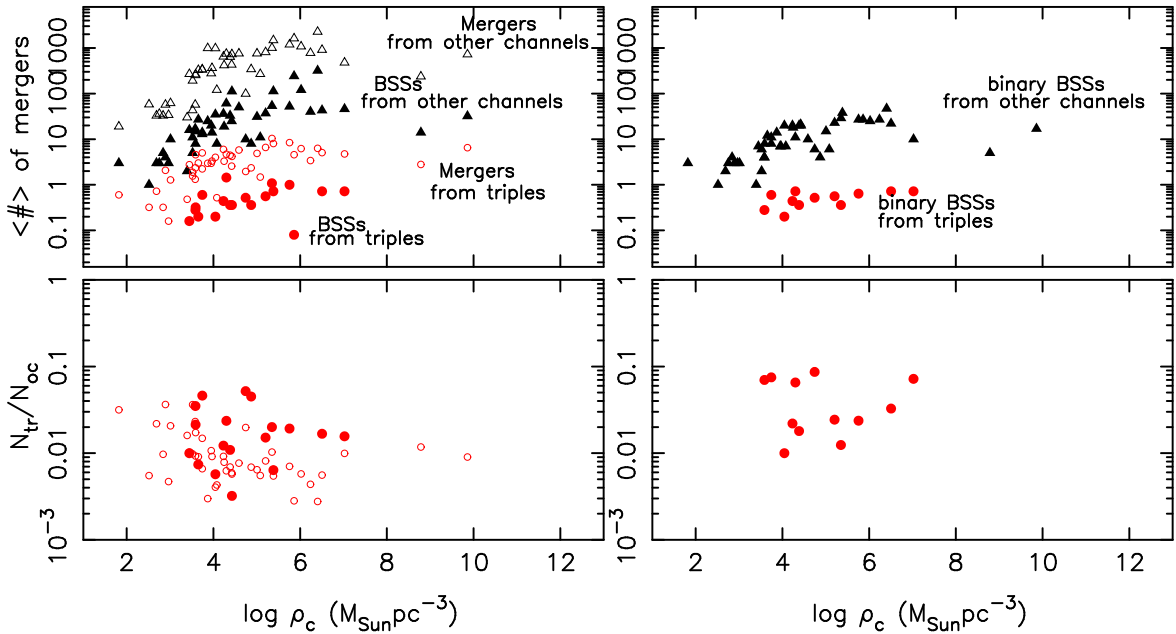

\centering
\includegraphics[width=3.3in,angle=270.]{Fig11a.eps}\includegraphics[width=3.3in,angle=270.]{Fig11b.eps}
\caption{ Contribution of triples to the population of BSSs (left panels) and binary BSSs (right panels) in GCs.
In the upper left panels, filled and open  circles represent respectively 
the total number of BSS candidates and stellar mergers formed in
triples.
The filled and open  triangles  show respectively the total number of BSSs and stellar mergers
in our cluster models that were not produced in hierarchical triples. 
In the right panels filled circles represent BSSs from hard triples
with MS star tertiary. These latter are 
the systems that will survive as binaries up to the present epoch and could be observed as binary BSSs.
In the lower panels the filled circles give the ratio of the number of (binary) BSSs
formed in triples ($N_{\rm tr}$) to the number of BSSs due to other channels\ ($N_{\rm oc}$).
The open circles give the number of stellar mergers from triples devided by the total number of 
stellar mergers from binary mediated channels.
These quantities are all plotted against the central
density of the model at $12\ $Gyr. Note that a few clusters are in a state of core-collapse at this time,
being characterized by very high central densities ($\rho_c \gtrsim 10^8 {\rm M_{\odot}pc^{-3}}$).
}\label{Fig11}
\end{figure*}

Figure\ \ref{Fig11} shows the relative importance of BSS production via traditional binary-mediated channels 
and triples. Since the triples were not followed in the Monte Carlo models, BSSs created via LK cycles 
leading to mass transfer or merger between the inner binary can only be calculated for each triple produced in 
the models. For these cases we cannot directly consider the subsequent stellar evolution of the BSSs. For example, 
whether followed by a merger the BSS created via a triple-mediated channel later evolves off of the MS before 
the observation time of $\sim 12\,\rm{Gyr}$, cannot be directly determined in this setup. Indeed, in detailed models 
it had been found that BSSs can live in the MS for several billion years after formation. It is also expected that 
the median age since formation for the observed BSSs can vary depending on the dominant formation channel for the 
BSSs in a real cluster \citep[e.g.,][]{2013ApJ...777..106C}. The real age since formation for BSSs, even in these detailed 
models might not be accurate, since the residual rejuvenated lifetime for a real BSS depends intricately on the details of 
the degree of H-mixing in the core which in turn depends on the details of the interaction that produced it 
\citep[e.g.,][]{2001ApJ...548..323S,2002ApJ...568..939L,2009MNRAS.395.1822C,2013ApJ...777..106C}. Instead, we report two quantities for the triple-mediated 
BSSs- 1) The total number of mergers or Roche-filling architectures created from dynamically formed triples during 
the full evolution of the model clusters, and  2) the same, but for binaries with 
 a total mass $\geq 1.05\times m_{\rm{TO}}$ and created only within the last $4\,\rm{Gyr}$, an estimated upper limit of today's 
observed BSSs in typical Galactic GCs \citep{2013ApJ...777..106C}. We find that the relative importance of triple-mediated 
channels is low compared to binary-mediated channels for BSS production. 
 The importance of triple-mediated channels is $\lesssim 10\%$ 
 for the cluster models considered in this work.

There is a lot of interest in BSSs found in binaries. Often these systems are thought to be produced via stable 
mass transfer in a binary where the donor stays bound to the accretor. 
On the other hand, these may also be 
produced via triple-mediated channels where the inner binary merges and the outer binary is hard and remains 
unbroken \citep[e.g.,][]{2007ApJ...669.1298F,2009ApJ...697.1048P,2015ebss.book..251P,2015arXiv151004290G}. 
We find that the contribution from triple-mediated channels 
for  binary BSSs is also small being 
$\lesssim 10\%$ of the number of mergers due to binary-mediated processes. In addition, we find that the binary BSSs 
coming via binary-mediated channels are \emph{not} simply coming from mass transfer in a binary. Collisionally produced 
BSSs can frequently acquire a binary companion via binary-single scattering encounters. Since BSSs are more massive 
compared to typical cluster stars at late ages, initially single BSSs,  
created earlier via a merger or physical collision can frequently acquire a binary companion 
via exchange with another normal star in a binary through binary-single interactions. This is 
consistent with earlier work that investigated the detailed history and branching ratios of various 
binary-mediated formation channels for BSSs produced in similar models 
\citep{2013ApJ...777..106C}. 

\section{Discussion}\label{disc}

\subsection{Implications for GW detectors and Comparison to other sources of eccentric GW inspirals}

The first GW signal from a compact object binary is likely to be detected in the
coming years by Advanced ground-based laser interferometers\ \citep{2015CQGra..32k5012A,2015CQGra..32b4001A}.
Most GW searches adopt matched filtering techniques in which the detector signal 
is cross-correlated with banks of theoretical gravitational waveforms to enable detection
of the weak signals. Mainly due to computational limitation, such banks of theoretical waveforms are 
exclusively composed of circular binaries. 
However, there are many reasons  to extend the searches to also include eccentric inspirals.
First, there have been recent developments of dedicated techniques that 
will potentially enable the 
detection of eccentric sources\ \citep{2014PhRvD..90j3001T,2014PhRvD..90h4016H,2015PhRvD..91f3004C,2015PhRvD..91l4014A}.
Second, eccentric inspirals produce a GW signal which is
distinct from that of circularly inspiraling binaries,
and provides more insights on the strong-field dynamics\
 \citep{2014PhRvD..90j4010L,2015arXiv150605648L,2015PhRvD..91l4014A}. 
Finally, as discussed below, there are several dynamical mechanisms that can lead to the
formation of eccentric GW sources. The detection of such eccentric sources could provide with important information about
properties (e.g., triple/binary fraction, central densities) of compact object populations in GCs and galactic nuclei that 
will be otherwise inaccessible.

Several mechanisms have been proposed for the formation of eccentrically inspiraling compact binaries 
for high frequency gravitational
wave detectors. These include: (i) BH-BH scattering in  steep density cusps around massive 
black holes\ \citep[MBHs;][]{2009MNRAS.395.2127O,2012PhRvD..85l3005K}, (ii) binary-single compact object encounters in star 
clusters\ \citep{2014ApJ...784...71S} and
(iii) LK resonance in hierarchical triple systems forming in the dense stellar environment of
galactic nuclei\ \citep{2012ApJ...757...27A} or stellar clusters\ 
\citep[e.g.,][this paper]{2003ApJ...598..419W}.
In this section we discuss the rate
for eccentric compact binary coalescences which 
could be detectable by aLIGO.
With the caution that rate estimates remain 
subject to significant uncertainties, we find that
BH binary mergers from hierarchical triples in GCs
are likely to  dominate the production  
of eccentric inspirals that could be detectable  by 
Advanced ground-based laser interferometers.

{\bf Single-single captures.}
In galactic nuclei with MBHs, nuclear relaxation times can be often less
than a Hubble time, and can result in the formation of steep density
cusps of  stellar-mass BHs as a consequence of mass segregation against the lower 
mass stars~\citep[e.g.,][]{2006ApJ...645L.133H}. 
\citet{2009MNRAS.395.2127O} proposed that in  such dense population environments
BH binaries can efficiently form out of gravitational wave emission during BH-BH encounters.
After forming, these BH binaries rapidly inspiral and merge
within a few hours. \citet{2009MNRAS.395.2127O} found that $\sim 90~\%$ of these coalescing binaries 
will have  an eccentricity larger than $0.9$ inside the aLIGO frequency band.
The predicted  rate for eccentric capture of $10M_{\odot}$ BH-BH encounters is
$\sim 0.01\rm yr^{-1} Gpc^{-3}$ \citep{2013ApJ...777..103T}. 

Table~1 of \citet{2009MNRAS.395.2127O}
presents the merger rate per MW-like galaxy
and for different cusp models.
The plausible pessimistic, likely, and optimistic rates
can be taken from models A$\beta 3$, E-2, and F-1
of their Table~1.
The most optimistic scenario, given by
model F-1, produces a merger rate of $1.5\times10^{-2}\rm \ Myr^{-1}$. The
realistic scenario, identified here with model  E-2, predicts $1.3\times10^{-3}\rm \ Myr^{-1}$.
The pessimistic scenario, given by model A$\beta 3$, produces $2\times10^{-4}\rm \ Myr^{-1}$. 
We can compare these merger rates with those corresponding to eccentric inspirals from GCs.
After multiplying the merger rates in our Table\ 1 by a factor $200$ \citep[roughly the number 
of GCs in the MW;][]{2014ApJ...797..128H}, we see that the
realistic merger rate for \citet{2009MNRAS.395.2127O} scenario
appears to be about one order of magnitude smaller than the realistic merger rate of 
eccentric inspirals from GCs.

\citet{2015MNRAS.448..754H} carried out  $N$-body simulations of star clusters around MBHs
 to study the formation of BH binaries in nuclear clusters
and to put constraints on the aLIGO detection rates for the
\citet{2009MNRAS.395.2127O} model.  \citet{2015MNRAS.448..754H}
derived an aLIGO expected detection rate of $0.02-14\ \rm yr^{-1}$
depending on the maximum horizon distance assumed and the mass ratio
of MBH to the surrounding cluster. 
As also note by \citet{2015MNRAS.448..754H}, these rates are likely to be an
overestimate of the real event rate from BH-BH captures.  They
were in fact obtained by assuming that the mass fraction of
BHs is $44\%$ of the total cluster mass, which is only reasonable 
within about one-tenth of the MBH influence radius of a mass-segregated nuclear
cluster\ \citep{2006ApJ...645L.133H}.  Given that most of the captures
were found to occur near the half-mass radius of the nuclear cluster,
where mass segregation will not significantly alter the initial number
fraction of different mass species \citep{2014ApJ...794..106A}, a more reasonable choice would be
to simply adopt a  BH mass fraction similar to that expected on the basis of the 
 initial mass function of the underlining population.  
For example, in the Galactic center there is evidence for a 
standard Kroupa-like initial mass function  \citep{2010MNRAS.402..519L}
which will result in $1\%$ only of the  total mass in BHs\ \citep{2006ApJ...645L.133H}. 
Using this standard choice for the initial mass function 
and adopting the same quadratic scaling on the BH mass fraction of \citet{2015MNRAS.448..754H}
will cause the estimated detection rates to drop by approximately 3 orders of magnitude.

{\bf Binary-single encounters.}
The in-cluster detection rate reported in Table\ 1  includes BH mergers due 
to single-binary interactions. Assuming that $1\%$ of these mergers will have
a finite eccentricity within the aLIGO band\ \citep{2014ApJ...784...71S}
results in a negligible contribution of binary-single encounters to the the population
of eccentric inspirals  when compared to the triple channel.

\citet{2014ApJ...784...71S} studied binary-single stellar scattering occurring in dense
star clusters as a source of eccentric NS-NS
inspirals for aLIGO. During the chaotic gravitational 
interaction between the three bodies, a pair of compact objects can be driven
to very high eccentricities such that they can inspiral through gravitational wave radiation and merge.
 \citet{2014ApJ...784...71S}  argued that $1\%$ of  dynamically assembled NS-NS merging binaries 
will have a substantial eccentricity at $\gtrsim 10\ {\rm Hz}$ frequency.
 \citet{2014ApJ...784...71S}  also give a simple order of magnitude estimate of the merger rate for dynamical 
NS-NS star eccentric inspirals assembled  in GCs of $0.7{\rm yr^{-1}Gpc^{-3}}$. 

To compare to the event rate we found for eccentric LK induced mergers in GCs we 
recompute the \citet{2014ApJ...784...71S} rates by rescaling with 
the different GC number density used in Section~\ref{d-rates}. 
\citet{2014ApJ...784...71S}  
calculate  the  rate of eccentric binaries in the aLIGO band 
adopting  $\rho_{\rm GC}=10\ {\rm GC~Mpc^{-3}}$.
Taking $\rho_{\rm GC}=0.77\ {\rm GC~Mpc^{-3}}$ instead, assuming 
that aLIGO will see NS-NS mergers up to  a sky-averaged distance of
$400\ {\rm Mpc}$~\citep{2010CQGra..27q3001A}, and optimistically  assuming that mergers
are distributed uniformly over the lifetime of the globulars (as
opposed to happening early on in the lifetime of the cluster),
this translates into a possible aLIGO detection rate for NS-NS eccentric inspirals of 
$\sim 0.02 {\rm yr^{-1}}$.

We note that \citet{2015PhRvD..91f3004C} give in their Table\ II
the potential detection rates 
for the model  of \citet{2014ApJ...784...71S}. These rates appear to be larger
by one order of magnitude  than our rate estimate above.
 We believe however that the rates computed
by \citet{2015PhRvD..91f3004C} were mistakenly enhanced due to 
a misinterpretation of \citet{2014ApJ...784...71S} results.
\citet{2014ApJ...784...71S} found that about ``$1\%$ of \emph{dynamically} assembled non-eccentric binaries''
will have a finite eccentricity at $10\ \rm Hz$.  \citet{2015PhRvD..91f3004C} computed the rates 
in their Table\ II as the  $1\%$ of the \emph{total} compact merger rate given in Table\ V of \citet{2010CQGra..27q3001A}
which  corresponds, however, to predictions for \emph{field} mergers.
In order to correct for this we have to multiply the rates in Table\ II of \citet{2015PhRvD..91f3004C} 
by the expected fraction of all compact object mergers that are dynamically assembled  in star clusters.
Taking this fraction to be $10\%$ of all NS-NS mergers\ \citep{2006NatPh...2..116G,2014ApJ...784...71S}, 
and assuming that $1\%$ of these are eccentric inspirals results in $4\times 10^{-3}{\rm yr^{-1}}$,
$0.04{\rm yr^{-1}}$, and $0.4{\rm yr^{-1}}$ for the low, realistic, and high detection rates 
for binary NSs.
These estimates appear to be in good agreement with the rate of $\sim 0.02 {\rm yr^{-1}}$ we previously computed.

 Although very simplified, the simple calculations presented above show that, 
even when assuming a uniform rate of mergers, 
the  rate of eccentric NS binaries in the aLIGO frequency band 
 is likely  1-2 orders of magnitude smaller than the realistic estimates for eccentric BH binaries
forming through the LK mechanism in GCs.

{\bf Massive black hole mediated compact-object mergers in galactic nuclei.}
\citet{2012ApJ...757...27A} studied the evolution of compact object binaries
orbiting a MBH.
Near the galactic center, where orbits are nearly Keplerian, binaries
form a triple system with the MBH in which
this latter represents the outer perturber of the triple. 
The MBH can subsequently drive the inner binary  orbit into high eccentricities
through LK resonance, at which point the compact object binary
can efficiently coalesce through GW emission.
\citet{2012ApJ...757...27A} found that $\approx 10\%$
of such coalescing binaries will  evolve through the dynamical non-secular evolution
similar to that described in Section~\ref{bd-oaa}, leading to
finite eccentricity sources in  the aLIGO sensitivity window.

\citet{2012ApJ...757...27A} estimated a
BH-BH eccentric inspiral  rate for a MW-like galaxy 
between $10^{-5}{\rm \ Myr^{-1}}$
and $10^{-2}{\rm \ Myr^{-1}}$. 
These rates are somewhat comparable to the rates for BH-BH mergers due to GW captures
derived by  \citet{2009MNRAS.395.2127O};
however, they are still  smaller than
the rates for 
eccentric inspirals assembled in GCs for a MW-like galaxy given in Table\ 1, 
and, even  under the most optimistic assumptions, they are still only
 $3 \%$ of the realistic 
and $0.01 \%$ of the high/optimistic rate in ~\citet{2010CQGra..27q3001A}.

 Given the realistic BH-BH detection rate
 with aLIGO of  $\sim 20{\rm yr^{-1}}$\ \citep{2010CQGra..27q3001A},
the predicted overall rate of eccentric compact binary coalescence in the aLIGO band  induced by MBH 
mediated LK cycles in galactic nuclei is $\sim 0.02{\rm yr^{-1}}$\ \citep{2015PhRvD..91f3004C}. 
We conclude that,
the overall contribution of eccentric GW sources from galactic nuclei is
likely  negligible compared to the rate of eccentric mergers arising from 
hierarchical triples in GCs.

\subsection{Implications for blue straggler formation}

The BSSs observed abundantly in all clusters have long been identified as a way 
of identifying important stages in the host cluster's history. However, interpretation 
of observed trends between the BSS number and cluster properties is 
dependent on the relative importance of the various BSS formation channels 
in these clusters. Especially, there has been a long history of carefully considering 
correlations between the number of BSSs produced in a cluster and several 
dynamically important observed cluster properties, including the total mass, total number of 
binaries, and the central collision/interaction rate, and what they mean
\citep{2004ApJ...604L.109P,2007ApJ...661..210L,2009Natur.457..288K,2013ApJ...777..106C}. 
Although, it has been pointed out that 
formation of BSSs from triple-mediated interactions may be important, at least in open clusters\ \citep{2011MNRAS.410.2370L}, 
this has never been self-consistently investigated in realistic and evolving cluster 
models \citep{2015ebss.book..251P}. As part of this study we have considered evolution of triples created dynamically 
in the cluster models and analyzed them as potential BSS progenitors. Upon comparison of the 
estimated numbers of the BSSs resulting from triple-mediated formation channels to those formed 
via binary-mediated channels (mass transfer and physical collisions due to binary-mediated interactions), 
we find that, at least for GC-like densities, contribution from triple-mediated channels is quite low. 

For the triple-mediated BSSs, we did not consider whether upon formation the rejuvenated MS star 
will remain in the region expected of BSSs in a color-magnitude diagram. The rejuvenated lifetime 
of BSSs is model dependent and hard to estimate. For the binary-mediated BSSs we assume full-mixing 
of Hydrogen.  In case of triple-mediated BSSs, we simply compare the total number of BSSs created throughout 
the lifetime of the GC and that formed within the last $4\,\rm{Gyr}$ and with mass 
$\geq 1.05\times m_{\rm{TO}}$. Thus, the actual contribution from triple-mediated 
interaction may be even lower if many of these BSSs evolve off of their rejuvenated MS before the observation 
time of $\sim 12\,\rm{Gyr}$. 

We note that above we have not considered 
BSS formation through binary evolution in short period binaries affected by the LK mechanism
coupled with tidal friction.  The shrinkage of the inner binary semi-major axis as
well as the high eccentricities induced via the LK mechanism, combined with
an expansion in the size of each inner binary star, could
lead to coalescing/strongly-interacting post-MS binaries.
 In order to address the importance of this alternative channel we 
determined whether the stars would undergo mass transfer by requiring
their periapsis distance to be smaller than the Roche limit 
computed through Equation~(\ref{roche}) and setting the 
stellar radius of the most massive component equal to that
evaluated at the tip of the red giant branch\ \citep{2000MNRAS.315..543H}.
The periapsis distance was evaluated using the orbital parameters at the end of the
three-body integrations. We only considered stars with mass $\gtrsim 0.7\ M_{\odot}$
so that their MS life time was shorter than the host GC age.
Also, we did not consider systems that will have undergone mass transfer 
on the red giant branch even without the help of LK cycles and tidal dissipation.
We found that only 694 of the total 19073 binaries that did not merge during the MS
would merge on the red giant branch through a combination of LK cycles, tidal friction
and the expansion of the stellar radius caused by stellar evolution. 
This calculation suggests that
the number of BSSs formed through a combination of 
LK dynamics and stellar evolution in our models is negligible.

Finally, note that we did not consider 
primordial triples  
that could somewhat increase the number of
mergers from stellar triples  shown in Figure\ \ref{Fig11}.
However, we think that mergers occurring in primordial triples will have a negligible contribution
to BSSs compared to mergers from dynamically assembled triples.
In fact, due to the high stellar densities of GCs 
the outer binaries in primordial triples are likely
to be involved in several dynamical encounters which will lead to their 
disruption in the first few Gyrs of cluster evolution.
The majority of triples in the catalog of \citet{2015ApJ...799....4R} have outer period $P_2\gtrsim 10^4\ $days,
which according to Equation~(\ref{coll}) corresponds to a typical survival time $T_{\rm enc}\lesssim 10^9\ $yr 
within the core of a GC and \ $T_{\rm enc}\lesssim 10^{10} $ in 
the cluster outskirts\ \citep[see also Fig. 2 of][]{2009ApJ...697.1048P}.
Moreover, given the old age of GCs
stellar merger products that occurred in primordial triples
would have already evolved off the main sequence, 
so that only  dynamically formed triples could then produce
BSSs at later stages of cluster evolution.

\section{Summary}\label{summ}
In this paper we studied the long-term evolution of dynamically assembled triples
in GCs. The initial conditions for the triples were obtained directly from
detailed Monte Carlo models of the clusters.
These systems consisted mostly of  triples containing an 
inner stellar binary and triples containing a pair of BHs.
The triple initial conditions  were integrated forward in time using a high
accuracy three-body integrator which incorporates relativistic corrections and tidal terms 
to the equations of motion. The direct integrations allowed us, for the first time, to 
put constraints on the contribution of the LK  
mechanism to the population of coalescing BH binaries and BSSs in GCs.
The main results of our study are briefly summarized below.

(1) We find that for the majority of dynamically assembled triple systems in GCs,
the time-scale for changes of angular momentum during the high eccentricity phase of a LK cycle
becomes shorter than the orbital time-scales.
We conclude that the standard orbit average equations of motion,
often employed in previous studies, cannot accurately describe the evolution of 
such triples. 

(2) BH triples assembled dynamically in GCs 
often evolve such that the inner binary angular momentum changes on timescales
of order or shorter than the inner binary orbital period (see Figure\ \ref{examples}). This can lead to
the formation of eccentric GW sources in the frequency band of aLIGO detectors\ (see Figure\ \ref{Fig3}).

(3) We estimated a realistic aLIGO detection rate of BH binary mergers 
due to the LK mechanism of $\sim 1\ \rm yr^{-1}$.
We also compared this rate to that for BH mergers due to 
other processes\ (see Table\ 1).
We find that mergers induced by LK oscillations are likely to be the main
contributor to in-cluster GW sources for
aLIGO detectors, and represent $\sim 1\%$ of the 
total detection rate of BH binary mergers assembled in
GCs.

(4) About $20\%$ of all LK mechanism induced BH mergers  in GCs
will have an eccentricity larger than $\sim 0.1$ at $10\ \rm Hz$ frequency,
 such that eccentric waveform templates and/or dedicated search 
strategies will be needed for their efficient detection (see Figure\ \ref{Fig4}). 
Most of these eccentric sources will have an extremely large eccentricity\ ($1-e\lesssim 10^{-4}$)
at the moment they first enter the aLIGO frequency band.
Assuming perfect waveform templates, we find a possible detection
rate for eccentric mergers as large as a few events a year 
with aLIGO. We also show that the triple channel is
likely the dominant formation mechanism for the production of eccentric inspirals 
that will be potentially detectable by aLIGO.

(5) In the case of triples containing a stellar binary, the inner binary angular momentum 
often changes on timescales shorter than the tertiary orbital period, but rarely 
shorter than the \emph{inner} binary orbital period. 
This implies that ``clean'' collisions between  two stars,
in which all passages prior to the collision are greater than the tidal dissipation scale, are 
rare (see Figure\ \ref{period-pdf}). 

(6) We estimated the number of BSSs expected to form through mergers induced by the LK mechanism
as well as those formed through other processes, including  binary stellar evolution, binary-single scattering and 
direct collisions\ (see Figure\ \ref{Fig11}).
We find that the contribution of LK induced mergers to the population of BSS populations in GCs 
is typically $\lesssim 10\%$ of the total. Even in clusters with relatively 
low central densities ($\sim 10^{2-3}\rm M_\odot{\rm pc^{-3}}$) only up to $\sim 10\%$ of BSS binaries could 
have formed in dynamically assembled triples.

\bigskip
We are  grateful to S.\ Mikkola who wrote
the ARCHAIN algorithm and who generously assisted 
us in using it. 
During the course of this work, we have benefited from
conversations with several colleagues, including Cole Miller and
Hagai Perets. We thank the referee for useful comments that helped to improve 
this paper. FA acknowledges support from a CIERA postdoctoral fellowship at Northwestern University. 
This work was supported by NASA ATP Grant NNX14AP92G and NSF Grant AST‐1312945. 
SC also acknowledges support from NASA through a grant from the 
Space Telescope Science Institute, which is operated by the 
Association of Universities for Research in Astronomy, Inc., under NASA contract NAS5-26555; 
the grant identifying number is HST-AR-12829.004-A. VK and FAR also acknowledge
support from NSF Grant PHY-1066293 through the Aspen Center for Physics.

  \end{document}